\documentclass[fleqn,usenatbib]{mnras}

\usepackage{newtxtext,newtxmath}

\usepackage[T1]{fontenc}

\DeclareRobustCommand{\VAN}[3]{#2}
\let\VANthebibliography\thebibliography
\def\thebibliography{\DeclareRobustCommand{\VAN}[3]{##3}\VANthebibliography}

\usepackage{graphicx}	
\usepackage{amsmath}	
\usepackage{xcolor}
\usepackage{multirow}
\usepackage{booktabs}  
\usepackage{lipsum}
\usepackage{pdflscape}
\usepackage{longtable}
\usepackage{booktabs}
\usepackage{geometry}

\title[AGC 322753]{AGC 322753: a Dark Galaxy Candidate detected by ALFALFA and FASHI}

\author[M. Monaci et al.]{Marco Monaci,$^{1}$\thanks{E-mail: mmonaci@swin.edu.au}
Duncan A. Forbes$^{1}$,
Warrick J. Couch$^{1}$,
and Jean P. Brodie$^{1}$
\\
$^{1}$Centre for Astrophysics and Supercomputing, Swinburne University of Technology, John Street, Hawthorn VIC 3122, Australia
}

\date{Accepted XXX. Received YYY; in original form ZZZ}

\pubyear{\the\year{}}

\begin{document}
\label{firstpage}
\pagerange{\pageref{firstpage}--\pageref{lastpage}}
\maketitle

\begin{abstract}
We report the possible detection of a Dark Galaxy Candidate (DGC; AGC 322753), via the cross-matching of the Arecibo Legacy Fast ALFA (ALFALFA) and the FAST All Sky \ion{H}{i} (FASHI) \ion{H}{i} blind surveys. It is detected by both surveys at the same recessional velocity, with a comparable \ion{H}{i} mass of $2 \times 10^{9} \ M_\odot$, and with similar \ion{H}{i} spectral features.
The ALFALFA and FASHI spectra show a double-peaked profile, compatible with a disc-like ordered rotation, and an \ion{H}{i} linewidth at 50 per cent of the peak (W$_{50}$) of $\sim85$ km s$^{-1}$.
The common intersection of the two beams allows the source position to be constrained with much greater precision than in either survey individually. Being detected by both radio telescopes, a Radio Frequency Interference (RFI) origin is excluded. No evidence of an optical counterpart was found down to $\sim$ 27.5 mag arcsec$^{-2}$ in the DESI Legacy Imaging Surveys, which leads to a stellar mass limit of $ M_\bigstar \lesssim 5 \times10^7 \ M_\odot$. Given this stellar mass limit, AGC 322753 has $M_{HI}/M_\bigstar \gtrsim 40$. 
The \ion{H}{i} mass is higher than expected for totally starless haloes; this could imply that AGC 322753 is not completely `dark' but could host a luminous counterpart below the current optical detection limit.
\end{abstract}



\section{Introduction}
The hierarchical mass assembly of cosmic structures is a key prediction of the $\Lambda$CDM model \citep[][]{1978MNRAS.183..341W}. In this framework, the most massive structures are considerably less numerous than very low-mass dark matter (DM) haloes, in which star formation (SF) is thought to be heavily suppressed \citep[][]{1992MNRAS.256P..43E, 2016MNRAS.457.1931S}. 

Several studies pointed out that there is a critical mass threshold for igniting the SF within a halo; however, its value is a matter of debate and depends on the redshift, the mass accretion history, the time when reionisation occurred, and the environment \citep[][]{2006MNRAS.371..401H, 2008MNRAS.390..920O, 2013MNRAS.432.3340S, 2020MNRAS.498.4887B, 2024MNRAS.529.3387A, 2026MNRAS.547ag385D}. Crucially, this threshold regulates the onset of SF, rather than its efficiency, which depends on local conditions, such as the local gas density, the dust-to-gas ratio, the far-UV flux, the metallicity, and turbulence \citep[][]{2004ApJ...609..667S, 2011ApJ...728...88G, 2016ApJ...826..200S}. Therefore, the study of these low-mass, `almost' or completely dark galaxies (DGs) could provide crucial insight into hierarchical galaxy formation, constrain the mechanisms suppressing SF, help constrain future cosmological simulations, and even help probe the physics of DM.

Early works \citep[][]{1997MNRAS.292L...5J, 2002MNRAS.336..541V, 2006MNRAS.368.1479D} generally agree that DGs have a low surface density that inhibits their SF and that they inhabit high-spin DM haloes, but should retain a fair amount of neutral hydrogen that, in principle, could be detected by radio surveys. \citet{2020MNRAS.498L..93J} revisited the model by \citet{1997MNRAS.292L...5J}, finding that (almost) dark galaxies are Toomre-stable since they reside in high-spin haloes, and that they have highly quenched SF rates, 3--4 orders of magnitude below that of the Milky Way. 

Over the past 15 years, the advent of large-scale hydrodynamical cosmological simulations has made it possible to investigate the properties of these objects in greater detail. \citet{2017MNRAS.465.3913B}, using the \texttt{APOSTLE} simulations, found a population of low-mass DM haloes with their gas in hydrostatic equilibrium with the DM gravitational potential and in thermal equilibrium with the UV background, which they named REionisation-Limited \ion{H}{i} Clouds (RELHICs). These objects have \ion{H}{i} masses below $3\times10^6 \ M_\odot$ and \ion{H}{i} linewidths dominated by thermal broadening ($W_{50}\lesssim 20 \rm \ km \ s^{-1}$), as RELHICs are not rotating. \citet{2024ApJ...962..129L}, using the TNG50 simulation suite, confirmed that dark galaxies preferentially reside in low-density environments, have larger spin parameters, and undergo fewer merger events.

Recently, \cite{2026ApJ..1004...79Z} investigated \ion{H}{i}-rich `almost' dark galaxies in the \texttt{HESTIA} and Auriga simulations \citep[][]{2020MNRAS.498.2968L, 2017MNRAS.467..179G} finding that their \ion{H}{i} masses overlap with that of the RELHICs, display low metallicity, but have a stellar component of $M_\bigstar\sim 4\times 10^5 \ M_{\odot}$ which could be difficult to detect in the optical. \cite{2026A&A...710A.156G} studied completely and `almost' ($M_\bigstar\lesssim 10^5 \ M_\odot$) DGs in \texttt{HESTIA} and \texttt{NIVARIA-LG}, finding that they reside in less-concentrated, higher-spin DM haloes than luminous galaxies; their mass distribution partially overlaps with that of the RELHICs, however, the most massive, starless galaxies, have \ion{H}{i} masses of 10$^7$ -- 10$^8 \ M_\odot$. Recently, \citet{2026arXiv260427047M} examined \ion{H}{i}- rich, starless haloes at $z = 0$ across three different cosmological simulations that successfully reproduce this kind of object.

In recent years, significant observational efforts have been made to identify starless haloes via their \ion{H}{i} emission. Cloud-9, discovered by \citet{2023ApJ...952..130Z} in the M94 group, and extensively studied by \citet{2023ApJ...956....1B}, \citet{2024ApJ...973...61B}, and \citet{2025ApJ...993L..55A}, is one of the best candidates for a RELHIC. It is dynamically cold ($W_{50}\sim 12 \rm \ km \ s^{-1}$) and has $M_{HI}\sim10^6 \ M_\odot$, without having an optical counterpart (OC) down to $M_\bigstar \lesssim 10^{4} \ M_\odot$.

From an observational point of view, we define a Dark Galaxy Candidate (DGC) as an extragalactic, isolated ($\sim 150 \rm \ kpc$ away from massive galaxies) \ion{H}{i} source without an OC in the deepest optical images available.
From this perspective, \citet{2025ApJS..279...38K} analysed the Arecibo Legacy Fast ALFA survey \citep[ALFALFA,][]{2018ApJ...861...49H}, finding 142 DGCs without a clear OC in the DESI Legacy Imaging Surveys \citep[hereafter Legacy Surveys,][]{2019AJ....157..168D}. However, they used a 75 kpc radius to exclude sources near other massive galaxies. Recently, \citet{2026MNRAS.548ag732M} analysed the DR1 of the FAST All Sky \ion{H}{i} survey \citep[FASHI,][]{2024SCPMA..6719511Z}, finding 70 DGCs within 50 Mpc from the Milky Way.

These objects are more massive in their \ion{H}{i} content than predicted for a RELHIC, and some of them show indications of ordered rotation. While it is observationally impossible to exclude a stellar body down to $M_\bigstar \sim 0$, deep optical images are crucial, as well as interferometric \ion{H}{i} observations. This is highlighted by the case of FAST J0139+4328. Initially claimed to be an isolated DG candidate with $M_{HI}\sim 8\times10^7 \ M_\odot$ by \citet{2023ApJ...944L..40X}, the OC was identified $\sim$0\farcm5 away from the \ion{H}{i} beam centroid through deep optical imaging by \citet{2026A&A...705L...9M}, who also spectroscopically confirmed its radial velocity (within $\sim$30 km s$^{-1}$ of the \ion{H}{i} source). Shortly after, the OC was also identified in a stacked Pan-STARRS image after pinpointing the \ion{H}{i} position with the Very Large Array \citep[][]{2026A&A...708A..40S}. The stellar mass estimates between the two works are fully consistent ($M_\bigstar \sim 5\times 10^6 \ M_\odot$), thereby reclassifying FAST J0139+4328 as a faint dwarf galaxy.

In the search for DGCs through \ion{H}{i} observations, Radio Frequency Interference (RFI) represents a serious problem with all radio telescopes. For example, the WALLABY \ion{H}{i} survey on ASKAP suffered from severe solar interference, forcing some data collected in the daytime to be excluded from analysis and future observations to only be scheduled at night. Therefore, the most reliable way to rule this out is to confirm the \ion{H}{i} source with two independent radio telescopes, using different receivers and located at different sites on Earth. Here, we present such a case, AGC 322753. It has been detected by both the ALFALFA and FASHI surveys, with compatible \ion{H}{i} properties and a double-peaked \ion{H}{i} profile indicative of a disc-like ordered rotation. The intersection between the two beams allows for improved precision on the position of a potential OC, and the depth of the Legacy Surveys allows a search for OCs down to $\sim$ 27.5 mag arcsec$^{-2}$ in the $g$ filter.

\section{Catalogues and selection process}\label{sec:cat_sel}
Here we use publicly available catalogues to search for DGCs in ALFALFA and FASHI \ion{H}{i} surveys, finding 13 sources in common. After a further check, we exclude 12 sources that, although interesting, are not the best candidates for a DGC (see the online Supplementary Material). In this section, we describe the catalogues used and the selection process to identify the one best example of a DGC.

\subsection{FAST All-Sky HI survey (FASHI)}\label{subsec:fashi}
The Five-hundred-metre Aperture Spherical Telescope \citep[FAST,][]{2011IJMPD..20..989N,2020RAA....20...64J} is the most powerful single-dish radio telescope in the world. Its generous aperture of 500 m reaches an unprecedented sensitivity while maintaining a beam size of $\sim3$\arcmin. Aside from other projects, FAST is being used to perform an \ion{H}{i} blind survey of the entire observable sky, with a velocity resolution of 6.4 km s$^{-1}$ and a median detection sensitivity of 0.76 mJy beam$^{-1}$. \cite{2024SCPMA..6719511Z} published the first release of the FASHI catalogue, containing 41741 extragalactic \ion{H}{i} sources up to redshift $z\simeq0.09$, and it partially overlaps with the ALFALFA survey within 30\degr < Dec < 36\degr. Low-reliability sources with a spectral S/N < 5 were excluded and are not listed in the final catalogue. The final catalogue lists several properties of each \ion{H}{i} source, such as the heliocentric velocity ($cz_{\odot}^{\rm FASHI}$), the linewidth at 50 and 20 per cent of the peak flux density (W$_{50}^{\rm FASHI}$, W$_{20}^{\rm FASHI}$), the inferred distance and the \ion{H}{i} mass.

\subsection{The FASHI catalogue of DGCs}\label{subsec:fastNDark}
Using the FASHI first release, \citet{2026MNRAS.548ag732M} compiled a catalogue of DGCs. The selection procedure involved both automatic and manual steps, starting with a velocity filtering (sources with $cz_\odot \leq 3500 \ \rm km \ s^{-1}$), then cross-matching the \ion{H}{i} sources with known galaxies. About 1000 \ion{H}{i} sources remained without a cross-match, and these were checked using the DESI Legacy Imaging Surveys. After a search for sources without an OC within the 3\arcmin \ beam and situated at least 150 kpc away from any massive galaxies (to avoid tidal tails), they identified 70 DGCs.

\subsection{The Arecibo Legacy Fast ALFA (ALFALFA) survey}\label{subsec:alfalfa}
The ALFALFA survey \citep[][]{2018ApJ...861...49H} covers $\sim 7000$ deg$^{2}$ at high Galactic latitude, over two large patches on the sky, up to redshift $z\sim 0.06$. The native velocity resolution is 5.1 km s$^{-1}$, while after the Hanning smoothing, it is 10 km s$^{-1}$. The map rms is 1.86 mJy beam$^{-1}$. The beam size is slightly larger (and elliptical) than that of FAST, namely $3\farcm8 \times 3\farcm3$, and for the purpose of this work, we will consider it as a circular beam of $3\farcm 5$ in diameter. The final ALFALFA catalogue comprises 25434 sources classified as `high' quality with generally an S/N $\geq 6.5$. The ALFALFA survey lists several properties, such as the heliocentric velocity $cz_{\odot}^{\rm ALFA}$, the W$_{50}^{\rm ALFA}$ and W$_{20}^{\rm ALFA}$, the distance and the \ion{H}{i} mass.

\subsection{The ALFALFA catalogue of DGCs}
Starting from 344 ALFALFA \ion{H}{i} sources, \cite{2025ApJS..279...38K} catalogued 142 DGCs without associated OCs (which they refer to with the code `0'), up to $cz_\odot\simeq 17470 \rm \ km \ s^{-1}$ ($\sim 254 \rm \ Mpc$). Their selection excluded some sources that are probably tidal in origin, some that are probably associated with nearby galaxies, and others that probably have an OC. They validated their selection by inspecting the Legacy Surveys, GALEX, and infrared surveys to see if any faint OCs are present.

\subsection{Selection process}\label{subsec:selection_process}
We start by cross-matching the \cite{2025ApJS..279...38K} list of 142 DGCs from ALFALFA with the full FASHI catalogue (Section \ref{subsec:fashi}). The surveys are neither fully homogeneous nor complete even across the overlapping region, so we expect the number of common sources to be low. We use quite relaxed conditions as a first step, imposing a maximum distance between the FASHI and ALFALFA sources of 3\arcmin (the worst-case scenario would therefore be an overlap between the two beams of about 0\farcm25) and a maximum allowed velocity difference of $|\Delta \rm v| < 300$ km s$^{-1}$. We do not use any other conditions, such as compatibility of \ion{H}{i} masses or W$_{50}$. After this cross-match, we obtain 13 potential matches, which are not included in the DGC catalogue compiled by \citet{2026MNRAS.548ag732M} because they all have $cz_\odot > 3500 \rm \ km \ s^{-1}$, and therefore were excluded by the velocity filtering.

The matches are robust, since all the sources agree within a few tens of km s$^{-1}$. Regarding the positions between the ALFALFA and FASHI detections of the same source, the beam centroids are all within 2\farcm2. After this first step, we proceed with the final evaluation of the sources, checking if they are near $(\lesssim 150 \ \rm kpc)$ other luminous galaxies or if there are potential OCs in the Legacy Surveys, and inspecting the ALFALFA \ion{H}{i} spectra. This process excluded 12 sources: 4 are located near bright galaxies, for 5, it was not possible to perform the visual check for various reasons (see online Supplementary Material for details), and 3 show potential OCs. AGC 322753 is the best candidate, since it shows a double-peaked profile in the ALFALFA \ion{H}{i} spectrum and there are no evident OCs within the two beams (see Figure \ref{fig:legacy_beams}). Here, the ALFALFA and FASHI beam centres are $\sim$2\farcm2 apart, sharing some overlap, which limits the search area for an OC.
The nearest galaxy, although it shares a similar radial velocity $(cz_\odot \sim 6910 \ \rm km \ s^{-1} )$, is  $\sim$ 290 kpc away in projection.

\section{Discussion}\label{sec:discussion}
In this section, we discuss the principal properties of AGC 322753 inferred from the \ion{H}{i} observations, checking whether the properties are consistent between the two detections, placing them in context with theoretical predictions, and estimating an upper limit for the stellar mass. 

\subsection{Presenting AGC 322753}\label{subsec:agc_presentation}
AGC 322753 was observed by the Arecibo telescope on behalf of the ALFALFA project. FASHI reported an \ion{H}{i} detection near the registered position of AGC 322753, centred about 2\farcm2 south-west of the ALFALFA detection. This detection is identified as J224556.66+335611.9 in FASHI; however, in this work, we use the AGC number for brevity. 

\begin{figure}
    \includegraphics[width=\columnwidth]{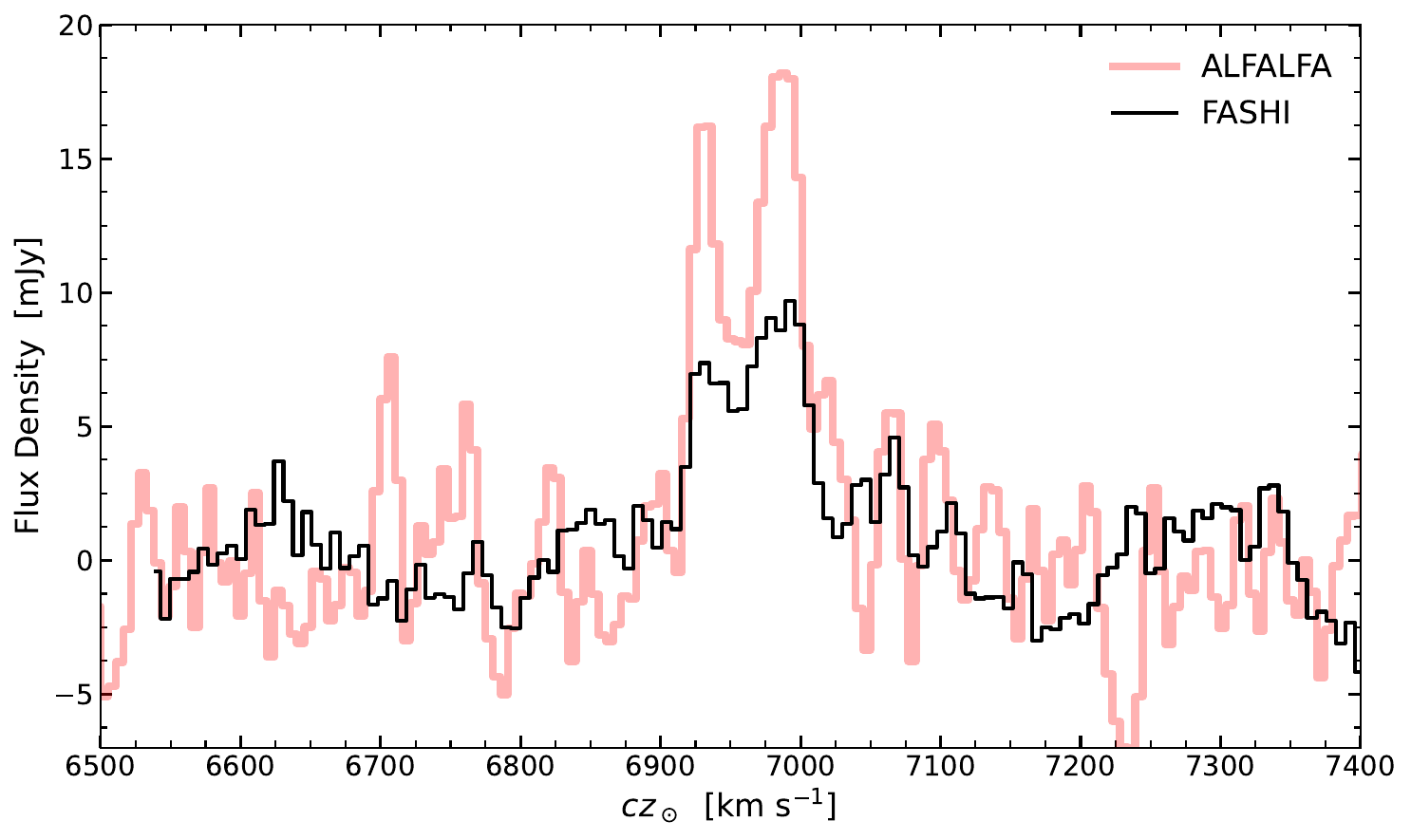}
    \caption{The \ion{H}{i} spectrum of AGC 322753 from the ALFALFA and FASHI surveys. The line shape, as well as the mean velocity, is in excellent agreement. Both spectra clearly show a double-peaked profile, indicative of rotation. The difference in the peak intensity is described in the text.}
    \label{fig:spectrum}
\end{figure}

We show in Figure \ref{fig:spectrum} the ALFALFA and FASHI \ion{H}{i} spectra of AGC 322753, which both show a clear double-peaked profile. The mean velocity and line shape are in excellent agreement between the two surveys. This does not hold for the peak flux density; it is a factor of 2 lower in the FASHI spectrum. \citet{2024SCPMA..6719511Z} pointed out a systematic in the integrated intensity for sources with an S/N $\lesssim 20$, that is, the FASHI integrated intensity is generally lower than the ALFALFA integrated intensity (see the bottom left panel of their figure 8). However, the \ion{H}{i} properties, as reported in Table \ref{tab:AGC_properties}, are in excellent agreement between the two surveys, except perhaps the \ion{H}{i} mass, which nevertheless depends on the integrated intensity. Given the same $cz_\odot$, W$_{50}$, and W$_{20}$, we conclude that these detections are of the same \ion{H}{i} source, observed by two independent radio telescopes at different epochs. This clearly excludes an RFI origin.

\begin{table*}
\centering
\caption{Comparison of various parameters as listed in ALFALFA and FASHI. ALFALFA does not directly list the errors on $cz_{\odot}$, so we used here the velocity resolution of 5 km s$^{-1}$. We note that the recently released FASHI DR2 \citep[][]{2026SCPMA..6929811Z} contains the same \ion{H}{i} source. However, the detection S/N is slightly lower, leading to a smaller \ion{H}{i} flux and mass, along with a slightly wider \ion{H}{i} profile. We suspect this may be the result of different sampling in the drift scanning process.}
\label{tab:AGC_properties}
\begin{tabular}{l|cccccccc}
\toprule
{\bf Survey} & Name & RA [deg] & Dec [deg] & $cz_{\odot}$ [km\,s$^{-1}$] & $W_{50}$ [km\,s$^{-1}$] & $W_{20}$ [km\,s$^{-1}$] & $\log_{10}(M_{\mathrm{HI}}/M_{\odot})$ & S/N \\
\midrule
{\bf ALFALFA} & AGC 322753           & 341.5150 & 33.9645 & $6963 \pm 5$        & $85 \pm 9$ & $110 \pm 9$      & $9.39 \pm 0.06$ & 11.8  \\
{\bf FASHI}   & J224556.66+335611.9 & 341.4861 & 33.9366 & $6967.19 \pm 1.12$ & $84.93 \pm 2.24$ & $100.77 \pm 3.35$ & $9.16 \pm 0.05$ & 19.1 \\
\bottomrule
\end{tabular}
\end{table*}

The most striking feature is the double-peaked profile of the \ion{H}{i} line, indicative of a rotating disc. In the case of a double-peaked profile, the W$_{50}$ can be used to infer the dynamical mass of the galaxy, since it traces an ordered rotation rather than random motions of the gas. However, without any information about the inclination of the disc, W$_{50}$ represents a lower limit for the rotational velocity. Assuming the simplest geometry of a disc seen edge-on, and considering an \ion{H}{i} radius of $\sim13 \ \rm kpc$ \citep[using the $R_{HI}-M_{HI}$ relation discussed by][]{2016MNRAS.460.2143W}, the inferred dynamical mass is $\sim 5.5\times10^9 \ M_\odot$, slightly higher than the \ion{H}{i} mass.
In order for AGC 322753 to accommodate the presence of DM, an inclination of $\ll 90^\circ$ is needed. Furthermore, an inclination of $\sim 30^\circ$ is needed for AGC 322753 to be consistent with the baryonic Tully–Fisher relation \citep[BTFR, see][]{2018MNRAS.474.4366P}.

\begin{figure}
    \includegraphics[width=\columnwidth]{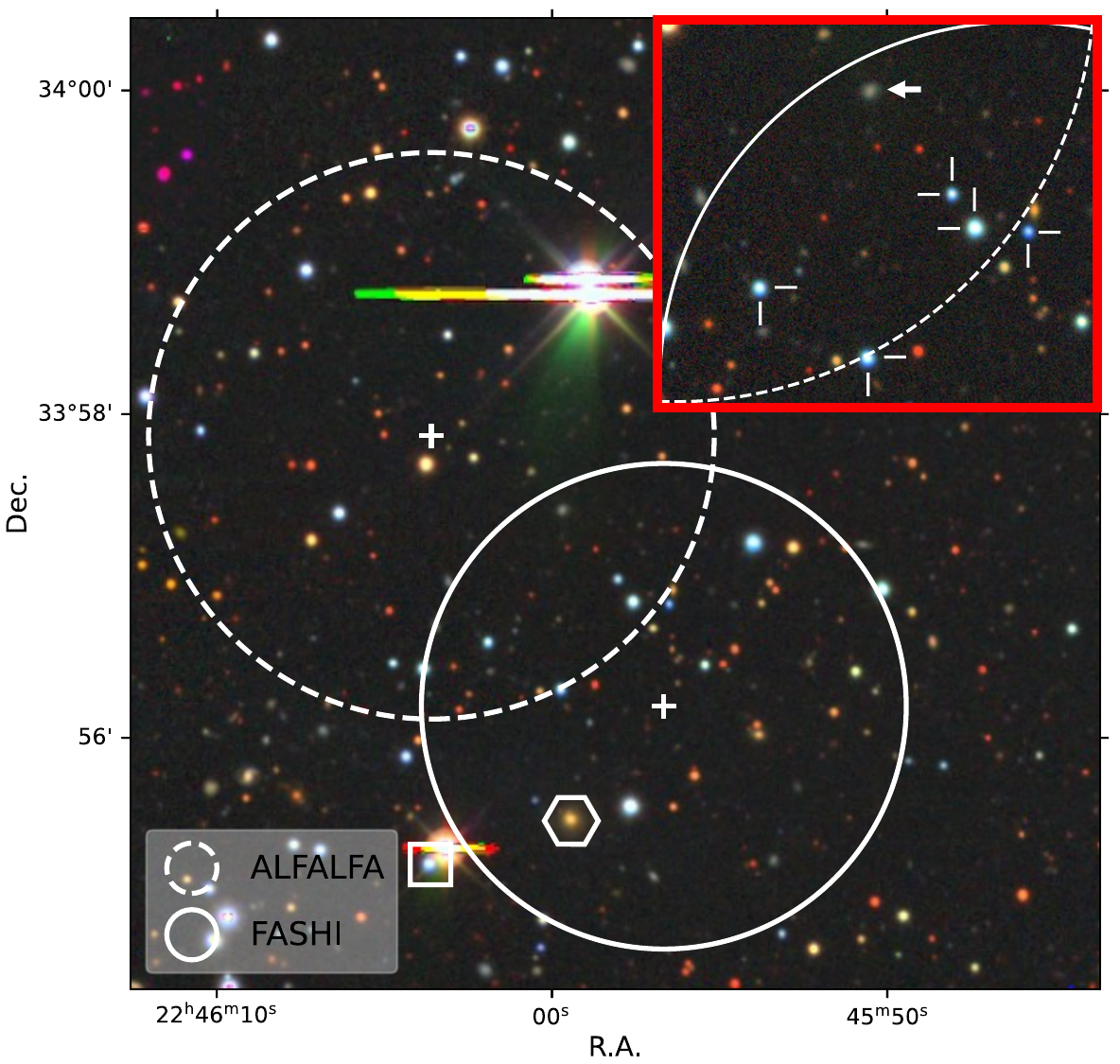}
    \caption{Colour image made by combining the $g$, $r$, and $i$ filters from Legacy Surveys. The dashed circle represents the 3\farcm5 beam of the Arecibo radio telescope, while the solid circle represents the 3\arcmin  ~beam of FAST. The crosses indicate the beam centroids. The white square highlights the position of the claimed OC by \citet{2024SCPMA..6719511Z} for this FASHI source. The white hexagon highlights the background galaxy (SDSS J224559+335532). The intersection between the two beams is the most probable location for the OC of AGC 322753, highlighted in the red inset. Here, marked sources with the short horizontal and vertical lines are foreground stars confirmed with Gaia DR3, and the galaxy marked by a white arrow is in the background. See the text.}
    \label{fig:legacy_beams}
\end{figure}

In Figure \ref{fig:legacy_beams}, we show the position on the sky of the ALFALFA and FASHI detections, with a dashed and a solid white circle, respectively. The crosses indicate the centroid positions as listed in the two surveys. The diameter of these circles (3\farcm5 for Arecibo and 3\arcmin~for FAST) represents the beam sizes. In the original ALFALFA catalogue, no OC is listed for this source.

\cite{2024SCPMA..6719511Z} performed a cross-match between FASHI and the SDSS photometric catalogue to search for a potential OC. They linked the FASHI detection of AGC 322753 with a potential OC, which we highlight with a white square in Figure \ref{fig:legacy_beams}. However, this potential OC is slightly off the 3\arcmin \ beam of the FAST telescope, and at first inspection of the Legacy Surveys, it appears to be a foreground star. Although SDSS classifies this object as a galaxy, it is included in Gaia DR3 \citep[][]{2023A&A...674A...1G}, where it is catalogued as a Milky Way star at a distance of approximately 1.9 kpc (photometric distance) and with a proper motion of $\sim 1.8 \rm \ mas \ yr^{-1}$. Furthermore, this source is well outside the ALFALFA beam, effectively ruling it out as a possible OC of AGC 322753. 

The only evident galaxy in the field of Figure \ref{fig:legacy_beams} is the one highlighted by a white hexagon within the FAST beam (SDSS J224559+335532). However, the photometric radial velocity from the Legacy Surveys of $cz_\odot=(6.9\pm 0.3)\times10^4 \rm \ km \ s^{-1}$ places it in the background. Finally, this galaxy is well outside the ALFALFA beam, rendering this source extremely unlikely to be the OC of AGC 322753.

\citet{2024SCPMA..6719511Z}, using the Siena Galaxy Atlas and the spectroscopic SDSS catalogue, reported that in the FASHI catalogue $\sim94$ per cent of the OCs are located within 1\farcm5 of the FAST beam centroid (and only $\sim$3 per cent at 2\arcmin \ from the beam centroid), so we expect that AGC 322753 is also located within 1\farcm5 from the FAST beam centre. The FAST beam partially overlaps with the Arecibo beam, and while we acknowledge that the \ion{H}{i} source could be outside both of the radio beams, this possibility is rather unlikely, and the shared region represents the most probable area where a potential OC could be present (see the inset in Figure \ref{fig:legacy_beams}). 
Within this intersection, we excluded some potential OCs. The sources marked with the short horizontal and vertical lines are foreground stars that have proper motions in Gaia DR3 (previously selected because they have a compatible photometric redshift). The small galaxy marked with a white arrow has a photometric $cz_\odot\simeq(8.5\pm1.5)\times10^4 \ \rm km \ s^{-1}$, implying it is in the background. We also checked the images from SDSS, Pan-STARRS, GALEX, and WISE, without finding any counterpart. As a final check, we downloaded and stacked the $gri$ images from the Legacy Surveys, without finding any faint OC. With no other evident candidates, AGC 322753 is a good Dark Galaxy Candidate.

Concerning the environment of AGC 322753, this galaxy is not completely isolated. The two nearest galaxies are UGC 12179 $(cz_\odot = 7005 \pm 29 \ \rm km \ s^{-1})$ and PGC 069656 $(cz_\odot = 6910 \pm 6 \ \rm km \ s^{-1})$, outside of Figure \ref{fig:legacy_beams}. They share almost the same recessional velocity as AGC 322753 $(cz_\odot = 6963 \pm 5 \ \rm km \ s^{-1})$. While we cannot exclude a tidal tail scenario in principle, where this object is an \ion{H}{i} cloud resulting from previous interactions between luminous galaxies in this group, several pieces of evidence argue against this interpretation. First, the double-peaked \ion{H}{i} profile is consistent with a rotating galaxy. Second, assuming a mean distance of 95 Mpc, UGC 12179 and PGC 069656 are located, in projection, 330 kpc and 290 kpc away, respectively. \citet{2020arXiv200207312K} pointed out that tails or plumes in interacting galactic systems typically extend up to 150 kpc and have \ion{H}{i} masses of up to $\sim 10^9$ M$_\odot$. If AGC 322753 were indeed a tidal feature, it would be one of the most massive (with $M_{HI}\sim 10^{9.3} \ M_\odot$) and extended tidal tails ever discovered.

\subsection{The W$_{50}$--M$_{\rm HI}$ plane}\label{subsec:W50-MHI}
\begin{figure}
   \includegraphics[width=\columnwidth]{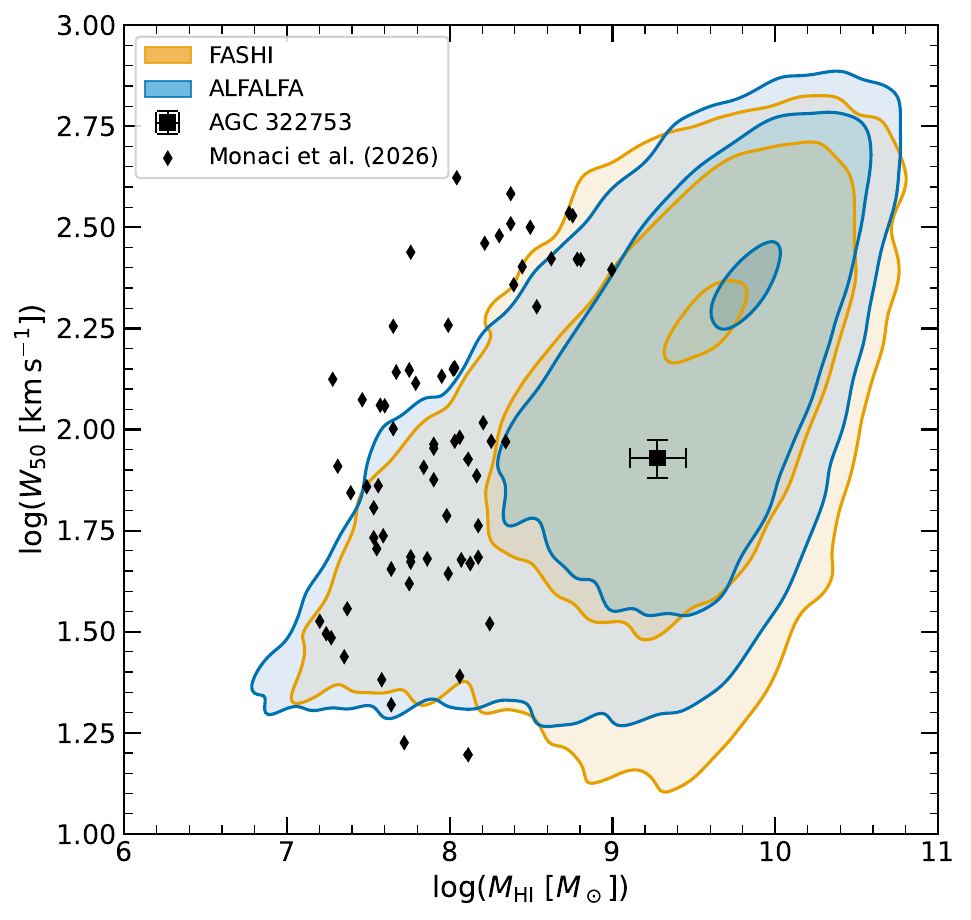}
    \caption{The position of AGC 322753 in the W$_{50}$--M$_{\rm HI}$ plane, compared with the full FASHI and ALFALFA samples. The contours for each distribution are at 10, 90, and 99 per cent, from the innermost to the outermost. The black diamonds are the DGCs from the catalogue compiled by \citet{2026MNRAS.548ag732M}. AGC 322753, although it has a consistent W$_{50}$ with these DGCs, has a higher \ion{H}{i} mass. The error bars are evaluated as described in the text.}
    \label{fig:MHI_W50_plane}
\end{figure}

The W$_{50}$--M$_{\rm HI}$ plane is a valuable tool to discuss the relation between AGC 322753 and other galaxies observed by ALFALFA and FASHI. Figure \ref{fig:MHI_W50_plane} shows the distribution of the full ALFALFA sample (blue contours) and FASHI sample (orange contours). The two samples are in good agreement, showing a linear trend between W$_{50}$ and the \ion{H}{i} mass. This trend is expected, as galaxies with larger \ion{H}{i} masses tend to reside in more massive haloes and therefore exhibit higher rotation velocities. The scatter is partly due to W$_{50}$ not being corrected for disc inclination. 

To calculate the position of AGC 322753, we use the mean for each quantity between the values listed in ALFALFA and FASHI. We used as the error for the \ion{H}{i} mass the maximum span of values from both surveys (specifically, taking $M_{\rm HI}-\Delta M_{\rm HI}$ from FASHI and $M_{\rm HI}+\Delta M_{\rm HI}$ from ALFALFA). For W$_{50}$, we used the errors listed in ALFALFA, since those reported by FASHI are significantly smaller.

As a comparison, in Figure \ref{fig:MHI_W50_plane} we also report the DGCs from the work by \citet{2026MNRAS.548ag732M}, remembering here that they are limited in distance to within 50 Mpc. AGC 322753, while showing a W$_{50}$ compatible with that of these DGCs, has an \ion{H}{i} mass that is larger than the masses of the DGCs within 50 Mpc. We further discuss this point and the implications for the stellar mass limit in Section \ref{subsec:stellar_limit}.

\subsection{Comparison with theoretical models}\label{subsec:theo_comp}
We now briefly review the pertinent theoretical works on dark and `almost' dark galaxies, and then we discuss AGC 322753 in light of these studies. That DM haloes can host stable gaseous reservoirs without forming stars was recognised by \citet{1986MNRAS.218P..25R} and \citet{1986Ap&SS.118..509I}, who proposed the scenario of photoionised gas in hydrostatic equilibrium within haloes deep enough to retain gas. Since AGC 322753 appears to be rotationally supported, it is incompatible with gaseous haloes in simple hydrostatic equilibrium.
Quantitative predictions for more massive and rotationally supported systems were later given by \citet{2006MNRAS.368.1479D}, who used simulated discs in NFW haloes to derive the expected \ion{H}{i} masses and $W_{50}$ values for dark galaxies.  In this regard, AGC 322753 is compatible with their $W_{50}$ predictions ($W_{50}\lesssim 100 \rm \ km \ s^{-1}$), but its mass is too high for it to be a completely dark galaxy in their scenario.

Given the \ion{H}{i} mass, $W_{50}$, and the double-peaked \ion{H}{i} profile, AGC 322753 is incompatible with the RELHIC picture \citep[][]{2017MNRAS.465.3913B}. RELHICs are non-rotating, with the \ion{H}{i} linewidths dominated by thermal broadening ($W_{50}\sim 20 \ \rm km \ s^{-1}$, considering a gas temperature of $T\sim10^4 \ K$). Their \ion{H}{i} masses extend up to $3\times 10^6 \ M_\odot$, therefore excluding AGC 322753.

\citet{2026ApJ..1004...79Z} investigated the \texttt{HESTIA} and \texttt{Auriga} simulations searching for \ion{H}{i}-rich, optically faint galaxies (referred to as HIDES). Interestingly, they did not find any completely starless haloes, with the exception of one. Generally, HIDES show $M_{HI}\sim3\times 10^6 \ M_\odot$ and $M_\bigstar \sim 4\times10^5 \ M_\odot$. Shortly after, \citet{2026A&A...710A.156G} analysed the cohort of starless haloes in three different \texttt{HESTIA} simulations and in the \texttt{NIVARIA-LG} simulation. They confirmed the existence of a class of objects with \ion{H}{i} masses below $\sim10^6 \ M_\odot$, fully consistent with RELHICs. Some of their starless Dark Galaxies extend up to $M_{HI} \sim 10^8 \ M_\odot$ and $M_{halo} \sim 9\times 10^9 \ M_\odot$ (see the bottom panel of their figure 3).

AGC 322753 is therefore too massive $(M_{HI}\simeq2 \times10^9 \ M_\odot)$ to be totally starless, but could have a stellar mass too low to be detectable in the Legacy Surveys. We conclude that the \ion{H}{i} mass of AGC 322753 exceeds any current prediction of a completely starless Dark Galaxy, pointing out that deeper optical imaging, as well as interferometric \ion{H}{i} observations, will be crucial to assess the nature of this object.

\subsection{Stellar mass limit}\label{subsec:stellar_limit}
Obtaining a reliable stellar mass limit for an optically faint (or even undetected) galaxy detected in \ion{H}{i} is subject to substantial uncertainties. When only unresolved \ion{H}{i} observations are available, several assumptions are required, with the galaxy inclination being one of the most significant sources of uncertainty. Other factors such as the presence of stars, diffraction spikes, image defects, and variable noise all complicate any analysis. With this caveat in mind, we carried out different injections of mock galaxies directly into the Legacy Surveys $g$-band images.

The surface brightness of the mock galaxies depends on several parameters. We explored 25 different combinations of $R_e$ and $M_\bigstar$, using two different $M/L_g$ ratios, leaving the other parameters fixed. We considered a tilted exponential disc (S\'ersic index $n=1$, inclination $i = 60\degr$) assuming an intrinsic axial ratio $q_0 = 0.20$. For the photometry, we considered $M/L_g = 1$ and $M/L_g = 0.5 \ M_\odot/L_\odot$, obtained rescaling for the $g$-band the $M/L_B=0.65 \ M_\odot/L_\odot$ obtained by \citet{2026A&A...705L...9M} for FAST J0139+4328. We did not correct for the cosmological dimming, but we corrected for the Galactic extinction. Before the injection, we convolved the galaxy with a Gaussian PSF with $FWHM = 1.3\arcsec$. To estimate $R_e$, we scaled the $R_{HI}-M_{HI}$ relation \citep[][]{2016MNRAS.460.2143W}, since AGC 322753 does not have any evident OC. The $R_e/R_{HI}$ ratio is not constant, but we can reasonably assume a range between 0.15 and 0.4 \citep[see, for example,][]{2023MNRAS.519.1098C}. Considering an \ion{H}{i} mass of $2\times10^9 \ M_\odot$, AGC 322753 should have an $R_{HI} \simeq 13 \ \rm kpc$. Scaling $R_{HI}$ to $R_e$, a reasonable estimate is between $\sim 2$ kpc and $\sim 6$ kpc. We note here that the `dark' galaxy discovered by \citet{2023ApJ...944L..40X} has an $R_e/R_{HI} = 0.28$. We explored five different $M_\bigstar$ ($10^8, 7\times10^7,5\times10^7,2\times10^7,10^7$ $M_\odot$) and five different $R_e$ (2, 3, 4, 5, 6 kpc) for a total of 25 mock galaxies per M/L ratio. 

\begin{figure*}
   \includegraphics[width=\textwidth]{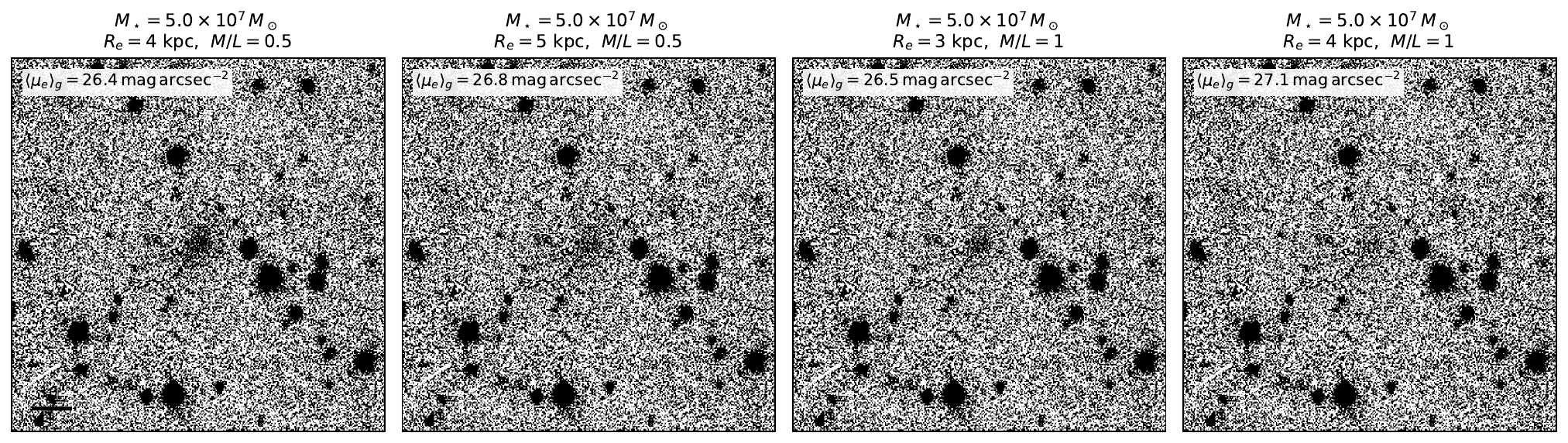}
    \caption{Four selected cases from the full set of injection runs available in the online Supplementary Material. The injection position is within the region where the two radio beams overlap, which is not affected by the stellar glare. The combination of parameters used for each injection is reported at the top of the corresponding panel, together with the mean surface brightness of the injected galaxy.}
    \label{fig:mock_selection}
\end{figure*}

In the $M/L_g = 0.5$ case, the $M_\bigstar = 5 \times 10^7 \ M_\odot$ galaxy is visible up to $R_e = 5 \ \rm kpc$, with the case $R_e = 6 \ \rm kpc$ at the limit. The $M_\bigstar = 2 \times 10^7 \ M_\odot$ galaxy is visible only if $R_e \leq 3 \rm \ kpc$. In the $M/L_g = 1$ case, the $M_\bigstar = 5 \times 10^7 \ M_\odot$ galaxy is visible if $R_e \leq 4 \ \rm kpc$, with the $R_e=4 \ \rm kpc$ case at the visibility limit. In the worst-case scenario, i.e. injection into the stellar glare, and with an $M/L_g = 1$, the galaxy is clearly visible if $M_\bigstar \sim 10^8 \ M_\odot$.  Furthermore, if the galaxy is face-on or has a spherical exponential profile, this limit would be even higher. Conversely, if the galaxy is disky, it is more edge-on, and we stack the $gri$ Legacy Surveys images, the limit would be lower. Leaving this analysis to future studies, we can reasonably infer a stellar mass limit of $M_\bigstar \lesssim 5\times 10^7 \ M_\odot$. In Figure \ref{fig:mock_selection} we show a selection of the limit cases. The full injections are available as online Supplementary Material.

Using this $M_\bigstar$ value, we can infer the $M_{HI}/M_\bigstar$. If we consider $5\times 10^7 \ M_\odot$ as a stellar mass limit, the $M_{HI}/M_\bigstar$ would be $\gtrsim 40$. To put this value in context, \citet{2026A&A...705L...9M} found a value of $11.5\pm 6.4$ for the `almost' dark galaxy FAST J0139+4328 discovered by \citet{2023ApJ...944L..40X}.

Future deep optical observations could help in finding a more stringent limit on the stellar mass, and with halo mass estimations, it will be possible to place AGC 322753 on the Stellar Mass-Halo Mass Relation (SHMR) and help constrain theoretical models in this halo mass regime where the scatter \citep[e.g.,][]{2020A&A...634A.135G} between models is significant.

\section{Conclusions}
In this Letter, we presented AGC 322753 as a Dark Galaxy Candidate (DGC) commonly found via \ion{H}{i} emission in both the ALFALFA and FASHI surveys, as conducted by the two most powerful single-dish radio telescopes, namely Arecibo and FAST. 

The \ion{H}{i} source found by both telescopes has the same recessional velocity within uncertainties, with similar \ion{H}{i} masses and inferred rotation from the double-peaked profile. The linewidth at 50 per cent of the peak (W$_{50}$) is about 85 km s$^{-1}$, in agreement between the two surveys. The derived \ion{H}{i} mass is also consistent across both datasets, with a value of $M_{HI} \simeq 2\times 10^9 \ M_\odot$. Both are strong detections, with S/N > 10, and we conclude they are the same \ion{H}{i} source.

The beams of the two surveys partially overlap, and their intersection represents the most probable location of the source. Consequently, even though we rely solely on single-dish surveys, the positional accuracy for locating the galaxy is significantly improved compared to each survey individually. Within this intersection region, the DESI Legacy Imaging Surveys show no evident optical counterpart (OC). Likewise, images from SDSS, Pan-STARRS, GALEX, and WISE do not reveal any clear indication of an OC.

The independent detection by two different radio telescopes, with different detectors, located in different places on Earth, and at different epochs, excludes an RFI origin and any other false positives that may be present in other studies.

Although we cannot rule out a tidal cloud origin, we note that the nearest luminous galaxies are at least 290 kpc away in projection. The significant rotation and high \ion{H}{i} mass also argue against a tidal cloud origin.

Regarding the consistency with theoretical predictions of completely starless haloes, AGC 322753 is incompatible with the RELHIC framework, given its large \ion{H}{i} mass and indications of rotational support.

Based on our injection of mock galaxies into Legacy $g$-band images, we estimated an upper limit on the stellar mass of $5 \times 10^7 \ \rm M_{\odot}$. With this constraint, AGC 322753 has $M_{\rm HI}/M_\bigstar \gtrsim 40$. We acknowledge that a more detailed approach is warranted for future work.   
Constraining both the stellar and halo mass would help to put AGC 322753 on the Stellar Mass-Halo Mass Relation (SHMR), in the halo mass regime where the SHMR shows a significant scatter.

In summary, AGC 322753 stands out as one of the most interesting and most promising DGCs in the ALFALFA and FASHI surveys. As such, it deserves follow-up studies, for example, obtaining spatially resolved \ion{H}{i} observations (e.g., with the VLA) to create a 2D rotation map, as well as deeper optical imaging (across multiple bands) to search for an extremely faint stellar counterpart.

\section*{Acknowledgements}
{\small We thank the anonymous Referee for the insightful comments that helped us
improve the original manuscript.}

{\small We thank Arianna Di Cintio, Guacimara Garc\'ia-Bethencourt, Kenji Bekki, and Anna Ivleva for fruitful discussions that contributed to shaping this study.}

{\small We thank Chuan-Peng Zhang and the FASHI collaboration for sharing the FASHI spectrum of AGC 322753.}

{\small MM acknowledges financial support received through a Swinburne University Postgraduate Research Award throughout the making of this work.}

{\small We thank the Australian Research Council (ARC) for financial support through the Discovery Project DP250101673.}

{\small This work has used the data from the Five-hundred-meter Aperture Spherical radio Telescope (FAST). FAST is a Chinese national mega-science facility, operated by the National Astronomical Observatories of Chinese Academy of Sciences (NAOC).}

{\small This research has made use of the NASA/IPAC Extragalactic Database (NED),
which is operated by the Jet Propulsion Laboratory, California Institute of Technology,
under contract with the National Aeronautics and Space Administration.}

{\small This research has made use of the VizieR catalogue access tool, CDS, Strasbourg, France.}

{\small We acknowledge the usage of the HyperLeda database (http://leda.univ-lyon1.fr).}

{\small The Photometric Redshifts for the Legacy Surveys (PRLS) catalog used in this paper was produced thanks to funding from the U.S. Department of Energy Office of Science, Office of High Energy Physics via grant DE-SC0007914.}

{\small The Legacy Surveys consist of three individual and complementary projects: the Dark Energy Camera Legacy Survey (DECaLS; Proposal ID \#2014B-0404; PIs: David Schlegel and Arjun Dey), the Beijing-Arizona Sky Survey (BASS; NOAO Prop. ID \#2015A-0801; PIs: Zhou Xu and Xiaohui Fan), and the Mayall z-band Legacy Survey (MzLS; Prop. ID \#2016A-0453; PI: Arjun Dey). DECaLS, BASS and MzLS together include data obtained, respectively, at the Blanco telescope, Cerro Tololo Inter-American Observatory, NSF’s NOIRLab; the Bok telescope, Steward Observatory, University of Arizona; and the Mayall telescope, Kitt Peak National Observatory, NOIRLab. Pipeline processing and analyses of the data were supported by NOIRLab and the Lawrence Berkeley National Laboratory (LBNL). The Legacy Surveys project is honored to be permitted to conduct astronomical research on Iolkam Du’ag (Kitt Peak), a mountain with particular significance to the Tohono O’odham Nation.
NOIRLab is operated by the Association of Universities for Research in Astronomy (AURA) under a cooperative agreement with the National Science Foundation. LBNL is managed by the Regents of the University of California under contract to the U.S. Department of Energy.
This project used data obtained with the Dark Energy Camera (DECam), which was constructed by the Dark Energy Survey (DES) collaboration. Funding for the DES Projects has been provided by the U.S. Department of Energy, the U.S. National Science Foundation, the Ministry of Science and Education of Spain, the Science and Technology Facilities Council of the United Kingdom, the Higher Education Funding Council for England, the National Center for Supercomputing Applications at the University of Illinois at Urbana-Champaign, the Kavli Institute of Cosmological Physics at the University of Chicago, Center for Cosmology and Astro-Particle Physics at the Ohio State University, the Mitchell Institute for Fundamental Physics and Astronomy at Texas A\&M University, Financiadora de Estudos e Projetos, Fundacao Carlos Chagas Filho de Amparo, Financiadora de Estudos e Projetos, Fundacao Carlos Chagas Filho de Amparo a Pesquisa do Estado do Rio de Janeiro, Conselho Nacional de Desenvolvimento Cientifico e Tecnologico and the Ministerio da Ciencia, Tecnologia e Inovacao, the Deutsche Forschungsgemeinschaft and the Collaborating Institutions in the Dark Energy Survey. The Collaborating Institutions are Argonne National Laboratory, the University of California at Santa Cruz, the University of Cambridge, Centro de Investigaciones Energeticas, Medioambientales y Tecnologicas-Madrid, the University of Chicago, University College London, the DES-Brazil Consortium, the University of Edinburgh, the Eidgenossische Technische Hochschule (ETH) Zurich, Fermi National Accelerator Laboratory, the University of Illinois at Urbana-Champaign, the Institut de Ciencies de l’Espai (IEEC/CSIC), the Institut de Fisica d’Altes Energies, Lawrence Berkeley National Laboratory, the Ludwig Maximilians Universitat Munchen and the associated Excellence Cluster Universe, the University of Michigan, NSF’s NOIRLab, the University of Nottingham, the Ohio State University, the University of Pennsylvania, the University of Portsmouth, SLAC National Accelerator Laboratory, Stanford University, the University of Sussex, and Texas A\&M University.
BASS is a key project of the Telescope Access Program (TAP), which has been funded by the National Astronomical Observatories of China, the Chinese Academy of Sciences (the Strategic Priority Research Program “The Emergence of Cosmological Structures” Grant \# XDB09000000), and the Special Fund for Astronomy from the Ministry of Finance. The BASS is also supported by the External Cooperation Program of Chinese Academy of Sciences (Grant \# 114A11KYSB20160057), and Chinese National Natural Science Foundation (Grant \# 12120101003, \# 11433005).
The Legacy Survey team makes use of data products from the Near-Earth Object Wide-field Infrared Survey Explorer (NEOWISE), which is a project of the Jet Propulsion Laboratory/California Institute of Technology. NEOWISE is funded by the National Aeronautics and Space Administration.
The Legacy Surveys imaging of the DESI footprint is supported by the Director, Office of Science, Office of High Energy Physics of the U.S. Department of Energy under Contract No. DE-AC02-05CH1123, by the National Energy Research Scientific Computing Center, a DOE Office of Science User Facility under the same contract; and by the U.S. National Science Foundation, Division of Astronomical Sciences under Contract No. AST-0950945 to NOAO.}

\section*{Data Availability}
This study has been conducted using publicly available data. For the FASHI spectra, please refer to \citet{2024SCPMA..6719511Z}. We used data from Legacy Surveys \citep[][]{2019AJ....157..168D}, Gaia \citep[][]{2023A&A...674A...1G}, photo-z PRLS \citep[][]{2021MNRAS.501.3309Z}, Siena Galaxy Atlas \citep[][]{2023ApJS..269....3M}, SDSS \citep[][]{2000AJ....120.1579Y}, Pan-STARRS1 \citep[][]{2016arXiv161205560C}, GALEX \citep[][]{2005ApJ...619L...1M}, WISE \citep[][]{2010AJ....140.1868W}.



\bibliographystyle{mnras}
\bibliography{biblio} 

@ARTICLE{2006MNRAS.368.1479D,
       author = {{Davies}, J.~I. and {Disney}, M.~J. and {Minchin}, R.~F. and {Auld}, R. and {Smith}, R.},
        title = "{The existence and detection of optically dark galaxies by 21-cm surveys}",
      journal = {\mnras},
         year = 2006,
        month = may,
       volume = {368},
       number = {3},
        pages = {1479-1488},
          doi = {10.1111/j.1365-2966.2006.10247.x},
archivePrefix = {arXiv},
       eprint = {astro-ph/0609747},
 primaryClass = {astro-ph},
       adsurl = {https://ui.adsabs.harvard.edu/abs/2006MNRAS.368.1479D}
}

@ARTICLE{2024SCPMA..6719511Z,
       author = {{Zhang}, Chuan-Peng and {Zhu}, Ming and {Jiang}, Peng and {Cheng}, Cheng and {Wang}, Jing and {Wang}, Jie and {Xu}, Jin-Long and {Liu}, Xiao-Lan and {Yu}, Nai-Ping and {Qian}, Lei and {Yu}, Haiyang and {Ai}, Mei and {Jing}, Yingjie and {Xu}, Chen and {Liu}, Ziming and {Guan}, Xin and {Sun}, Chun and {Yang}, Qingliang and {Huang}, Menglin and {Hao}, Qiaoli and {FAST Collaboration}},
        title = "{The FAST all sky H I survey (FASHI): The first release of catalog}",
      journal = {Science China Physics, Mechanics, and Astronomy},
         year = 2024,
        month = jan,
       volume = {67},
       number = {1},
          eid = {219511},
        pages = {219511},
          doi = {10.1007/s11433-023-2219-7},
archivePrefix = {arXiv},
       eprint = {2312.06097},
 primaryClass = {astro-ph.GA},
       adsurl = {https://ui.adsabs.harvard.edu/abs/2024SCPMA..6719511Z}
}

@ARTICLE{2020RAA....20...64J,
       author = {{Jiang}, Peng and {Tang}, Ning-Yu and {Hou}, Li-Gang and {Liu}, Meng-Ting and {Kr{\v{c}}o}, Marko and {Qian}, Lei and {Sun}, Jing-Hai and {Ching}, Tao-Chung and {Liu}, Bin and {Duan}, Yan and {Yue}, You-Ling and {Gan}, Heng-Qian and {Yao}, Rui and {Li}, Hui and {Pan}, Gao-Feng and {Yu}, Dong-Jun and {Liu}, Hong-Fei and {Li}, Di and {Peng}, Bo and {Yan}, Jun and {FAST Collaboration}},
        title = "{The fundamental performance of FAST with 19-beam receiver at L band}",
      journal = {Research in Astronomy and Astrophysics},
         year = 2020,
        month = may,
       volume = {20},
       number = {5},
          eid = {064},
        pages = {064},
          doi = {10.1088/1674-4527/20/5/64},
archivePrefix = {arXiv},
       eprint = {2002.01786},
 primaryClass = {astro-ph.IM},
       adsurl = {https://ui.adsabs.harvard.edu/abs/2020RAA....20...64J}
}

@ARTICLE{2020A&A...634A.135G,
       author = {{Girelli}, G. and {Pozzetti}, L. and {Bolzonella}, M. and {Giocoli}, C. and {Marulli}, F. and {Baldi}, M.},
        title = "{The stellar-to-halo mass relation over the past 12 Gyr. I. Standard {\ensuremath{\Lambda}}CDM model}",
      journal = {\aap},
         year = 2020,
        month = feb,
       volume = {634},
          eid = {A135},
        pages = {A135},
          doi = {10.1051/0004-6361/201936329},
archivePrefix = {arXiv},
       eprint = {2001.02230},
 primaryClass = {astro-ph.CO},
       adsurl = {https://ui.adsabs.harvard.edu/abs/2020A&A...634A.135G}
}

@ARTICLE{2011IJMPD..20..989N,
       author = {{Nan}, Rendong and {Li}, Di and {Jin}, Chengjin and {Wang}, Qiming and {Zhu}, Lichun and {Zhu}, Wenbai and {Zhang}, Haiyan and {Yue}, Youling and {Qian}, Lei},
        title = "{The Five-Hundred Aperture Spherical Radio Telescope (fast) Project}",
      journal = {International Journal of Modern Physics D},
         year = 2011,
        month = jan,
       volume = {20},
       number = {6},
        pages = {989-1024},
          doi = {10.1142/S0218271811019335},
archivePrefix = {arXiv},
       eprint = {1105.3794},
 primaryClass = {astro-ph.IM},
       adsurl = {https://ui.adsabs.harvard.edu/abs/2011IJMPD..20..989N}
}

@ARTICLE{2018ApJ...861...49H,
       author = {{Haynes}, Martha P. and {Giovanelli}, Riccardo and {Kent}, Brian R. and {Adams}, Elizabeth A.~K. and {Balonek}, Thomas J. and {Craig}, David W. and {Fertig}, Derek and {Finn}, Rose and {Giovanardi}, Carlo and {Hallenbeck}, Gregory and {Hess}, Kelley M. and {Hoffman}, G. Lyle and {Huang}, Shan and {Jones}, Michael G. and {Koopmann}, Rebecca A. and {Kornreich}, David A. and {Leisman}, Lukas and {Miller}, Jeffrey and {Moorman}, Crystal and {O'Connor}, Jessica and {O'Donoghue}, Aileen and {Papastergis}, Emmanouil and {Troischt}, Parker and {Stark}, David and {Xiao}, Li},
        title = "{The Arecibo Legacy Fast ALFA Survey: The ALFALFA Extragalactic H I Source Catalog}",
      journal = {\apj},
         year = 2018,
        month = jul,
       volume = {861},
       number = {1},
          eid = {49},
        pages = {49},
          doi = {10.3847/1538-4357/aac956},
archivePrefix = {arXiv},
       eprint = {1805.11499},
 primaryClass = {astro-ph.GA},
       adsurl = {https://ui.adsabs.harvard.edu/abs/2018ApJ...861...49H}
}

@ARTICLE{2025ApJS..279...38K,
       author = {{Kwon}, Minseong and {Hwang}, Ho Seong and {Kent}, Brian R. and {Yoon}, Ilsang and {Lee}, Gain and {Yoon}, Hyein},
        title = "{Searching for Dark Galaxies with H I Detection from the Arecibo Legacy Fast ALFA (ALFALFA) Survey}",
      journal = {\apjs},
         year = 2025,
        month = aug,
       volume = {279},
       number = {2},
          eid = {38},
        pages = {38},
          doi = {10.3847/1538-4365/ade0b8},
archivePrefix = {arXiv},
       eprint = {2506.03678},
 primaryClass = {astro-ph.GA},
       adsurl = {https://ui.adsabs.harvard.edu/abs/2025ApJS..279...38K}
}

@ARTICLE{2016MNRAS.460.2143W,
       author = {{Wang}, Jing and {Koribalski}, B{\"a}rbel S. and {Serra}, Paolo and {van der Hulst}, Thijs and {Roychowdhury}, Sambit and {Kamphuis}, Peter and {Chengalur}, Jayaram N.},
        title = "{New lessons from the H I size-mass relation of galaxies}",
      journal = {\mnras},
         year = 2016,
        month = aug,
       volume = {460},
       number = {2},
        pages = {2143-2151},
          doi = {10.1093/mnras/stw1099},
archivePrefix = {arXiv},
       eprint = {1605.01489},
 primaryClass = {astro-ph.GA},
       adsurl = {https://ui.adsabs.harvard.edu/abs/2016MNRAS.460.2143W}
}

@ARTICLE{2020arXiv200207312K,
       author = {{Koribalski}, B.~S.},
        title = "{Hydrogen tails, plumes, clouds and filaments}",
      journal = {arXiv e-prints},
         year = 2020,
        month = feb,
          eid = {arXiv:2002.07312},
        pages = {arXiv:2002.07312},
          doi = {10.48550/arXiv.2002.07312},
archivePrefix = {arXiv},
       eprint = {2002.07312},
 primaryClass = {astro-ph.GA},
       adsurl = {https://ui.adsabs.harvard.edu/abs/2020arXiv200207312K}
}

@ARTICLE{2026A&A...710A.156G,
       author = {{Garc{\'\i}a-Bethencourt}, Guacimara and {Di Cintio}, Arianna and {Comer{\'o}n}, S{\'e}bastien and {Arjona-G{\'a}lvez}, Elena and {Contreras-Santos}, Ana and {Cardona-Barrero}, Salvador and {Brook}, Chris B.~A. and {Negri}, Andrea and {Libeskind}, Noam I. and {Knebe}, Alexander},
        title = "{H I-bearing dark galaxy predictions from constrained Local Group simulations: How many and where to find them}",
      journal = {\aap},
         year = 2026,
        month = jun,
       volume = {710},
          eid = {A156},
        pages = {A156},
          doi = {10.1051/0004-6361/202558801},
archivePrefix = {arXiv},
       eprint = {2601.04024},
 primaryClass = {astro-ph.GA},
       adsurl = {https://ui.adsabs.harvard.edu/abs/2026A&A...710A.156G}
}

@ARTICLE{2026A&A...705L...9M,
       author = {{Mitra{\v{s}}inovi{\'c}}, Ana and {Grozdanovi{\'c}}, Marko and {Lalovi{\'c}}, Ana and {Jovanovi{\'c}}, Milena and {B{\'\i}lek}, Michal and {Pavlov}, Nata{\v{s}}a and {Moiseev}, Alexei V. and {Oparin}, Dmitry V.},
        title = "{Discovery of a galaxy associated with the HI cloud FAST J0139+4328}",
      journal = {\aap},
         year = 2026,
        month = jan,
       volume = {705},
          eid = {L9},
        pages = {L9},
          doi = {10.1051/0004-6361/202558391},
archivePrefix = {arXiv},
       eprint = {2512.24924},
 primaryClass = {astro-ph.GA},
       adsurl = {https://ui.adsabs.harvard.edu/abs/2026A&A...705L...9M}
}

@ARTICLE{2018MNRAS.474.4366P,
       author = {{Ponomareva}, Anastasia A. and {Verheijen}, Marc A.~W. and {Papastergis}, Emmanouil and {Bosma}, Albert and {Peletier}, Reynier F.},
        title = "{From light to baryonic mass: the effect of the stellar mass-to-light ratio on the Baryonic Tully-Fisher relation}",
      journal = {\mnras},
         year = 2018,
        month = mar,
       volume = {474},
       number = {4},
        pages = {4366-4384},
          doi = {10.1093/mnras/stx3066},
archivePrefix = {arXiv},
       eprint = {1711.09112},
 primaryClass = {astro-ph.GA},
       adsurl = {https://ui.adsabs.harvard.edu/abs/2018MNRAS.474.4366P}
}

@ARTICLE{2023A&A...674A...1G,
       author = {{Gaia Collaboration} and {Vallenari}, A. and {Brown}, A.~G.~A. and {Prusti}, T. and {de Bruijne}, J.~H.~J. and {Arenou}, F. and {Babusiaux}, C. and {Biermann}, M. and {Creevey}, O.~L. and {Ducourant}, C. and {Evans}, D.~W. and {Eyer}, L. and {Guerra}, R. and {Hutton}, A. and {Jordi}, C. and {Klioner}, S.~A. and {Lammers}, U.~L. and {Lindegren}, L. and {Luri}, X. and {Mignard}, F. and {Panem}, C. and {Pourbaix}, D. and {Randich}, S. and {Sartoretti}, P. and {Soubiran}, C. and {Tanga}, P. and {Walton}, N.~A. and {Bailer-Jones}, C.~A.~L. and {Bastian}, U. and {Drimmel}, R. and {Jansen}, F. and {Katz}, D. and {Lattanzi}, M.~G. and {van Leeuwen}, F. and {Bakker}, J. and {Cacciari}, C. and {Casta{\~n}eda}, J. and {De Angeli}, F. and {Fabricius}, C. and {Fouesneau}, M. and {Fr{\'e}mat}, Y. and {Galluccio}, L. and {Guerrier}, A. and {Heiter}, U. and {Masana}, E. and {Messineo}, R. and {Mowlavi}, N. and {Nicolas}, C. and {Nienartowicz}, K. and {Pailler}, F. and {Panuzzo}, P. and {Riclet}, F. and {Roux}, W. and {Seabroke}, G.~M. and {Sordo}, R. and {Th{\'e}venin}, F. and {Gracia-Abril}, G. and {Portell}, J. and {Teyssier}, D. and {Altmann}, M. and {Andrae}, R. and {Audard}, M. and {Bellas-Velidis}, I. and {Benson}, K. and {Berthier}, J. and {Blomme}, R. and {Burgess}, P.~W. and {Busonero}, D. and {Busso}, G. and {C{\'a}novas}, H. and {Carry}, B. and {Cellino}, A. and {Cheek}, N. and {Clementini}, G. and {Damerdji}, Y. and {Davidson}, M. and {de Teodoro}, P. and {Nu{\~n}ez Campos}, M. and {Delchambre}, L. and {Dell'Oro}, A. and {Esquej}, P. and {Fern{\'a}ndez-Hern{\'a}ndez}, J. and {Fraile}, E. and {Garabato}, D. and {Garc{\'\i}a-Lario}, P. and {Gosset}, E. and {Haigron}, R. and {Halbwachs}, J.-L. and {Hambly}, N.~C. and {Harrison}, D.~L. and {Hern{\'a}ndez}, J. and {Hestroffer}, D. and {Hodgkin}, S.~T. and {Holl}, B. and {Jan{\ss}en}, K. and {Jevardat de Fombelle}, G. and {Jordan}, S. and {Krone-Martins}, A. and {Lanzafame}, A.~C. and {L{\"o}ffler}, W. and {Marchal}, O. and {Marrese}, P.~M. and {Moitinho}, A. and {Muinonen}, K. and {Osborne}, P. and {Pancino}, E. and {Pauwels}, T. and {Recio-Blanco}, A. and {Reyl{\'e}}, C. and {Riello}, M. and {Rimoldini}, L. and {Roegiers}, T. and {Rybizki}, J. and {Sarro}, L.~M. and {Siopis}, C. and {Smith}, M. and {Sozzetti}, A. and {Utrilla}, E. and {van Leeuwen}, M. and {Abbas}, U. and {{\'A}brah{\'a}m}, P. and {Abreu Aramburu}, A. and {Aerts}, C. and {Aguado}, J.~J. and {Ajaj}, M. and {Aldea-Montero}, F. and {Altavilla}, G. and {{\'A}lvarez}, M.~A. and {Alves}, J. and {Anders}, F. and {Anderson}, R.~I. and {Anglada Varela}, E. and {Antoja}, T. and {Baines}, D. and {Baker}, S.~G. and {Balaguer-N{\'u}{\~n}ez}, L. and {Balbinot}, E. and {Balog}, Z. and {Barache}, C. and {Barbato}, D. and {Barros}, M. and {Barstow}, M.~A. and {Bartolom{\'e}}, S. and {Bassilana}, J.-L. and {Bauchet}, N. and {Becciani}, U. and {Bellazzini}, M. and {Berihuete}, A. and {Bernet}, M. and {Bertone}, S. and {Bianchi}, L. and {Binnenfeld}, A. and {Blanco-Cuaresma}, S. and {Blazere}, A. and {Boch}, T. and {Bombrun}, A. and {Bossini}, D. and {Bouquillon}, S. and {Bragaglia}, A. and {Bramante}, L. and {Breedt}, E. and {Bressan}, A. and {Brouillet}, N. and {Brugaletta}, E. and {Bucciarelli}, B. and {Burlacu}, A. and {Butkevich}, A.~G. and {Buzzi}, R. and {Caffau}, E. and {Cancelliere}, R. and {Cantat-Gaudin}, T. and {Carballo}, R. and {Carlucci}, T. and {Carnerero}, M.~I. and {Carrasco}, J.~M. and {Casamiquela}, L. and {Castellani}, M. and {Castro-Ginard}, A. and {Chaoul}, L. and {Charlot}, P. and {Chemin}, L. and {Chiaramida}, V. and {Chiavassa}, A. and {Chornay}, N. and {Comoretto}, G. and {Contursi}, G. and {Cooper}, W.~J. and {Cornez}, T. and {Cowell}, S. and {Crifo}, F. and {Cropper}, M. and {Crosta}, M. and {Crowley}, C. and {Dafonte}, C. and {Dapergolas}, A. and {David}, M. and {David}, P. and {de Laverny}, P. and {De Luise}, F. and {De March}, R.},
        title = "{Gaia Data Release 3. Summary of the content and survey properties}",
      journal = {\aap},
         year = 2023,
        month = jun,
       volume = {674},
          eid = {A1},
        pages = {A1},
          doi = {10.1051/0004-6361/202243940},
archivePrefix = {arXiv},
       eprint = {2208.00211},
 primaryClass = {astro-ph.GA},
       adsurl = {https://ui.adsabs.harvard.edu/abs/2023A&A...674A...1G}
}

@ARTICLE{2021MNRAS.501.3309Z,
       author = {{Zhou}, Rongpu and {Newman}, Jeffrey A. and {Mao}, Yao-Yuan and {Meisner}, Aaron and {Moustakas}, John and {Myers}, Adam D. and {Prakash}, Abhishek and {Zentner}, Andrew R. and {Brooks}, David and {Duan}, Yutong and {Landriau}, Martin and {Levi}, Michael E. and {Prada}, Francisco and {Tarle}, Gregory},
        title = "{The clustering of DESI-like luminous red galaxies using photometric redshifts}",
      journal = {\mnras},
         year = 2021,
        month = mar,
       volume = {501},
       number = {3},
        pages = {3309-3331},
          doi = {10.1093/mnras/staa3764},
archivePrefix = {arXiv},
       eprint = {2001.06018},
 primaryClass = {astro-ph.CO},
       adsurl = {https://ui.adsabs.harvard.edu/abs/2021MNRAS.501.3309Z}
}

@ARTICLE{2026ApJ..1004...79Z,
       author = {{Zheng}, Haonan and {Jiang}, Fangzhou and {Liao}, Shihong and {Libeskind}, Noam I.},
        title = "{HIDES. I. The Population and Diversity of H I-rich Faint Dwarf Galaxies in the HESTIA and Auriga Simulations}",
      journal = {\apj},
         year = 2026,
        month = jun,
       volume = {1004},
       number = {1},
          eid = {79},
        pages = {79},
          doi = {10.3847/1538-4357/ae6b84},
archivePrefix = {arXiv},
       eprint = {2511.16726},
 primaryClass = {astro-ph.GA},
       adsurl = {https://ui.adsabs.harvard.edu/abs/2026ApJ..1004...79Z}
}

@ARTICLE{2024MNRAS.529.3387A,
       author = {{Ahvazi}, Niusha and {Benson}, Andrew and {Sales}, Laura V. and {Nadler}, Ethan O. and {Weerasooriya}, Sachi and {Du}, Xiaolong and {Bovill}, Mia Sauda},
        title = "{A comprehensive model for the formation and evolution of the faintest Milky Way dwarf satellites}",
      journal = {\mnras},
         year = 2024,
        month = apr,
       volume = {529},
       number = {4},
        pages = {3387-3407},
          doi = {10.1093/mnras/stae761},
archivePrefix = {arXiv},
       eprint = {2308.13599},
 primaryClass = {astro-ph.GA},
       adsurl = {https://ui.adsabs.harvard.edu/abs/2024MNRAS.529.3387A}
}

@ARTICLE{2013MNRAS.432.3340S,
       author = {{Sobacchi}, Emanuele and {Mesinger}, Andrei},
        title = "{How does radiative feedback from an ultraviolet background impact reionization?}",
      journal = {\mnras},
         year = 2013,
        month = jul,
       volume = {432},
       number = {4},
        pages = {3340-3348},
          doi = {10.1093/mnras/stt693},
archivePrefix = {arXiv},
       eprint = {1301.6781},
 primaryClass = {astro-ph.CO},
       adsurl = {https://ui.adsabs.harvard.edu/abs/2013MNRAS.432.3340S}
}

@ARTICLE{2004ApJ...609..667S,
       author = {{Schaye}, Joop},
        title = "{Star Formation Thresholds and Galaxy Edges: Why and Where}",
      journal = {\apj},
         year = 2004,
        month = jul,
       volume = {609},
       number = {2},
        pages = {667-682},
          doi = {10.1086/421232},
archivePrefix = {arXiv},
       eprint = {astro-ph/0205125},
 primaryClass = {astro-ph},
       adsurl = {https://ui.adsabs.harvard.edu/abs/2004ApJ...609..667S}
}

@ARTICLE{2020MNRAS.498L..93J,
       author = {{Jimenez}, Raul and {Heavens}, Alan F.},
        title = "{The distribution of dark galaxies and spin bias}",
      journal = {\mnras},
         year = 2020,
        month = nov,
       volume = {498},
       number = {1},
        pages = {L93-L97},
          doi = {10.1093/mnrasl/slaa135},
archivePrefix = {arXiv},
       eprint = {2005.11798},
 primaryClass = {astro-ph.GA},
       adsurl = {https://ui.adsabs.harvard.edu/abs/2020MNRAS.498L..93J}
}

@ARTICLE{2023ApJ...952..130Z,
       author = {{Zhou}, Ruilei and {Zhu}, Ming and {Yang}, Yanbin and {Yu}, Haiyang and {Yuan}, Lixia and {Jiang}, Peng and {Xi}, Wenzhe},
        title = "{FAST Reveals New Evidence for M94 as a Merger}",
      journal = {\apj},
         year = 2023,
        month = aug,
       volume = {952},
       number = {2},
          eid = {130},
        pages = {130},
          doi = {10.3847/1538-4357/acdcf5},
archivePrefix = {arXiv},
       eprint = {2306.05080},
 primaryClass = {astro-ph.GA},
       adsurl = {https://ui.adsabs.harvard.edu/abs/2023ApJ...952..130Z}
}

@ARTICLE{2023ApJ...956....1B,
       author = {{Ben{\'\i}tez-Llambay}, Alejandro and {Navarro}, Julio F.},
        title = "{Is a Recently Discovered H I Cloud near M94 a Starless Dark Matter Halo?}",
      journal = {\apj},
         year = 2023,
        month = oct,
       volume = {956},
       number = {1},
          eid = {1},
        pages = {1},
          doi = {10.3847/1538-4357/acf767},
archivePrefix = {arXiv},
       eprint = {2309.03253},
 primaryClass = {astro-ph.GA},
       adsurl = {https://ui.adsabs.harvard.edu/abs/2023ApJ...956....1B}
}

@ARTICLE{2024ApJ...973...61B,
       author = {{Ben{\'\i}tez-Llambay}, Alejandro and {Dutta}, Rajeshwari and {Fumagalli}, Michele and {Navarro}, Julio F.},
        title = "{Examining the Nature of the Starless Dark Matter Halo Candidate Cloud-9 with Very Large Array Observations}",
      journal = {\apj},
         year = 2024,
        month = sep,
       volume = {973},
       number = {1},
          eid = {61},
        pages = {61},
          doi = {10.3847/1538-4357/ad65d9},
       adsurl = {https://ui.adsabs.harvard.edu/abs/2024ApJ...973...61B}
}

@ARTICLE{2025ApJ...993L..55A,
       author = {{Anand}, Gagandeep S. and {Ben{\'\i}tez-Llambay}, Alejandro and {Beaton}, Rachael and {Fox}, Andrew J. and {Navarro}, Julio F. and {D'Onghia}, Elena},
        title = "{The First RELHIC? Cloud-9 is a Starless Gas Cloud}",
      journal = {\apjl},
         year = 2025,
        month = nov,
       volume = {993},
       number = {2},
          eid = {L55},
        pages = {L55},
          doi = {10.3847/2041-8213/ae1584},
archivePrefix = {arXiv},
       eprint = {2508.20157},
 primaryClass = {astro-ph.GA},
       adsurl = {https://ui.adsabs.harvard.edu/abs/2025ApJ...993L..55A}
}

@ARTICLE{2026MNRAS.547ag385D,
       author = {{Doppel}, Jessica E. and {Jauzac}, Mathilde and {Lagattuta}, David J. and {Fattahi}, Azadeh and {Mahler}, Guillaume},
        title = "{Tiny galaxies and dark substructures: exploring the 'dark' subhaloes in TNG50}",
      journal = {\mnras},
         year = 2026,
        month = apr,
       volume = {547},
       number = {3},
          eid = {stag385},
        pages = {stag385},
          doi = {10.1093/mnras/stag385},
archivePrefix = {arXiv},
       eprint = {2506.09122},
 primaryClass = {astro-ph.GA},
       adsurl = {https://ui.adsabs.harvard.edu/abs/2026MNRAS.547ag385D}
}

@ARTICLE{2011ApJ...728...88G,
       author = {{Gnedin}, Nickolay Y. and {Kravtsov}, Andrey V.},
        title = "{Environmental Dependence of the Kennicutt-Schmidt Relation in Galaxies}",
      journal = {\apj},
         year = 2011,
        month = feb,
       volume = {728},
       number = {2},
          eid = {88},
        pages = {88},
          doi = {10.1088/0004-637X/728/2/88},
archivePrefix = {arXiv},
       eprint = {1004.0003},
 primaryClass = {astro-ph.CO},
       adsurl = {https://ui.adsabs.harvard.edu/abs/2011ApJ...728...88G}
}

@ARTICLE{2016ApJ...826..200S,
       author = {{Semenov}, Vadim A. and {Kravtsov}, Andrey V. and {Gnedin}, Nickolay Y.},
        title = "{Nonuniversal Star Formation Efficiency in Turbulent ISM}",
      journal = {\apj},
         year = 2016,
        month = aug,
       volume = {826},
       number = {2},
          eid = {200},
        pages = {200},
          doi = {10.3847/0004-637X/826/2/200},
archivePrefix = {arXiv},
       eprint = {1512.03101},
 primaryClass = {astro-ph.GA},
       adsurl = {https://ui.adsabs.harvard.edu/abs/2016ApJ...826..200S}
}

@ARTICLE{2026SCPMA..6929811Z,
       author = {{Zhang}, Chuan-Peng and {Zhu}, Ming and {Jiang}, Peng and {Guo}, Hong and {Xu}, Jin-Long and {Liu}, Xiao-Lan and {Yu}, Nai-Ping and {Cheng}, Cheng and {Wang}, Jing and {Wang}, Jie and {FAST Collaboration}},
        title = "{The FAST all sky HI survey DR2: The FASHI catalog and the HI mass function}",
      journal = {Science China Physics, Mechanics, and Astronomy},
         year = 2026,
        month = sep,
       volume = {69},
       number = {12},
          eid = {129811},
        pages = {129811},
          doi = {10.1007/s11433-026-3072-1},
archivePrefix = {arXiv},
       eprint = {2606.31539},
 primaryClass = {astro-ph.GA},
       adsurl = {https://ui.adsabs.harvard.edu/abs/2026SCPMA..6929811Z}
}

@ARTICLE{1997MNRAS.292L...5J,
       author = {{Jimenez}, R. and {Heavens}, A.~F. and {Hawkins}, M.~R.~S. and {Padoan}, P.},
        title = "{Dark galaxies, spin bias and gravitational lenses}",
      journal = {\mnras},
         year = 1997,
        month = nov,
       volume = {292},
       number = {1},
        pages = {L5-L10},
          doi = {10.1093/mnras/292.1.L5},
archivePrefix = {arXiv},
       eprint = {astro-ph/9709050},
 primaryClass = {astro-ph},
       adsurl = {https://ui.adsabs.harvard.edu/abs/1997MNRAS.292L...5J}
}

@ARTICLE{2026MNRAS.548ag732M,
       author = {{Monaci}, Marco and {Forbes}, Duncan A. and {Gannon}, Jonah S. and {Koribalski}, B{\"a}rbel S. and {Bekki}, Kenji and {Brodie}, Jean P. and {Couch}, Warrick J.},
        title = "{FAST and Dark: a catalogue of dark galaxy candidates within 50 Mpc}",
      journal = {\mnras},
         year = 2026,
        month = may,
       volume = {548},
       number = {3},
          eid = {stag732},
        pages = {stag732},
          doi = {10.1093/mnras/stag732},
archivePrefix = {arXiv},
       eprint = {2604.14699},
 primaryClass = {astro-ph.GA},
       adsurl = {https://ui.adsabs.harvard.edu/abs/2026MNRAS.548ag732M}
}

@ARTICLE{2023ApJS..269....3M,
       author = {{Moustakas}, John and {Lang}, Dustin and {Dey}, Arjun and {Juneau}, St{\'e}phanie and {Meisner}, Aaron and {Myers}, Adam D. and {Schlafly}, Edward F. and {Schlegel}, David J. and {Valdes}, Francisco and {Weaver}, Benjamin A. and {Zhou}, Rongpu},
        title = "{Siena Galaxy Atlas 2020}",
      journal = {\apjs},
         year = 2023,
        month = nov,
       volume = {269},
       number = {1},
          eid = {3},
        pages = {3},
          doi = {10.3847/1538-4365/acfaa2},
archivePrefix = {arXiv},
       eprint = {2307.04888},
 primaryClass = {astro-ph.GA},
       adsurl = {https://ui.adsabs.harvard.edu/abs/2023ApJS..269....3M}
}

@ARTICLE{2000AJ....120.1579Y,
       author = {{York}, Donald G. and {Adelman}, J. and {Anderson}, Jr., John E. and {Anderson}, Scott F. and {Annis}, James and {Bahcall}, Neta A. and {Bakken}, J.~A. and {Barkhouser}, Robert and {Bastian}, Steven and {Berman}, Eileen and {Boroski}, William N. and {Bracker}, Steve and {Briegel}, Charlie and {Briggs}, John W. and {Brinkmann}, J. and {Brunner}, Robert and {Burles}, Scott and {Carey}, Larry and {Carr}, Michael A. and {Castander}, Francisco J. and {Chen}, Bing and {Colestock}, Patrick L. and {Connolly}, A.~J. and {Crocker}, J.~H. and {Csabai}, Istv{\'a}n and {Czarapata}, Paul C. and {Davis}, John Eric and {Doi}, Mamoru and {Dombeck}, Tom and {Eisenstein}, Daniel and {Ellman}, Nancy and {Elms}, Brian R. and {Evans}, Michael L. and {Fan}, Xiaohui and {Federwitz}, Glenn R. and {Fiscelli}, Larry and {Friedman}, Scott and {Frieman}, Joshua A. and {Fukugita}, Masataka and {Gillespie}, Bruce and {Gunn}, James E. and {Gurbani}, Vijay K. and {de Haas}, Ernst and {Haldeman}, Merle and {Harris}, Frederick H. and {Hayes}, J. and {Heckman}, Timothy M. and {Hennessy}, G.~S. and {Hindsley}, Robert B. and {Holm}, Scott and {Holmgren}, Donald J. and {Huang}, Chi-hao and {Hull}, Charles and {Husby}, Don and {Ichikawa}, Shin-Ichi and {Ichikawa}, Takashi and {Ivezi{\'c}}, {\v{Z}}eljko and {Kent}, Stephen and {Kim}, Rita S.~J. and {Kinney}, E. and {Klaene}, Mark and {Kleinman}, A.~N. and {Kleinman}, S. and {Knapp}, G.~R. and {Korienek}, John and {Kron}, Richard G. and {Kunszt}, Peter Z. and {Lamb}, D.~Q. and {Lee}, B. and {Leger}, R. French and {Limmongkol}, Siriluk and {Lindenmeyer}, Carl and {Long}, Daniel C. and {Loomis}, Craig and {Loveday}, Jon and {Lucinio}, Rich and {Lupton}, Robert H. and {MacKinnon}, Bryan and {Mannery}, Edward J. and {Mantsch}, P.~M. and {Margon}, Bruce and {McGehee}, Peregrine and {McKay}, Timothy A. and {Meiksin}, Avery and {Merelli}, Aronne and {Monet}, David G. and {Munn}, Jeffrey A. and {Narayanan}, Vijay K. and {Nash}, Thomas and {Neilsen}, Eric and {Neswold}, Rich and {Newberg}, Heidi Jo and {Nichol}, R.~C. and {Nicinski}, Tom and {Nonino}, Mario and {Okada}, Norio and {Okamura}, Sadanori and {Ostriker}, Jeremiah P. and {Owen}, Russell and {Pauls}, A. George and {Peoples}, John and {Peterson}, R.~L. and {Petravick}, Donald and {Pier}, Jeffrey R. and {Pope}, Adrian and {Pordes}, Ruth and {Prosapio}, Angela and {Rechenmacher}, Ron and {Quinn}, Thomas R. and {Richards}, Gordon T. and {Richmond}, Michael W. and {Rivetta}, Claudio H. and {Rockosi}, Constance M. and {Ruthmansdorfer}, Kurt and {Sandford}, Dale and {Schlegel}, David J. and {Schneider}, Donald P. and {Sekiguchi}, Maki and {Sergey}, Gary and {Shimasaku}, Kazuhiro and {Siegmund}, Walter A. and {Smee}, Stephen and {Smith}, J. Allyn and {Snedden}, S. and {Stone}, R. and {Stoughton}, Chris and {Strauss}, Michael A. and {Stubbs}, Christopher and {SubbaRao}, Mark and {Szalay}, Alexander S. and {Szapudi}, Istvan and {Szokoly}, Gyula P. and {Thakar}, Anirudda R. and {Tremonti}, Christy and {Tucker}, Douglas L. and {Uomoto}, Alan and {Vanden Berk}, Dan and {Vogeley}, Michael S. and {Waddell}, Patrick and {Wang}, Shu-i. and {Watanabe}, Masaru and {Weinberg}, David H. and {Yanny}, Brian and {Yasuda}, Naoki and {SDSS Collaboration}},
        title = "{The Sloan Digital Sky Survey: Technical Summary}",
      journal = {\aj},
         year = 2000,
        month = sep,
       volume = {120},
       number = {3},
        pages = {1579-1587},
          doi = {10.1086/301513},
archivePrefix = {arXiv},
       eprint = {astro-ph/0006396},
 primaryClass = {astro-ph},
       adsurl = {https://ui.adsabs.harvard.edu/abs/2000AJ....120.1579Y}
}

@ARTICLE{2016arXiv161205560C,
       author = {{Chambers}, K.~C. and {Magnier}, E.~A. and {Metcalfe}, N. and {Flewelling}, H.~A. and {Huber}, M.~E. and {Waters}, C.~Z. and {Denneau}, L. and {Draper}, P.~W. and {Farrow}, D. and {Finkbeiner}, D.~P. and {Holmberg}, C. and {Koppenhoefer}, J. and {Price}, P.~A. and {Rest}, A. and {Saglia}, R.~P. and {Schlafly}, E.~F. and {Smartt}, S.~J. and {Sweeney}, W. and {Wainscoat}, R.~J. and {Burgett}, W.~S. and {Chastel}, S. and {Grav}, T. and {Heasley}, J.~N. and {Hodapp}, K.~W. and {Jedicke}, R. and {Kaiser}, N. and {Kudritzki}, R.-P. and {Luppino}, G.~A. and {Lupton}, R.~H. and {Monet}, D.~G. and {Morgan}, J.~S. and {Onaka}, P.~M. and {Shiao}, B. and {Stubbs}, C.~W. and {Tonry}, J.~L. and {White}, R. and {Ba{\~n}ados}, E. and {Bell}, E.~F. and {Bender}, R. and {Bernard}, E.~J. and {Boegner}, M. and {Boffi}, F. and {Botticella}, M.~T. and {Calamida}, A. and {Casertano}, S. and {Chen}, W.-P. and {Chen}, X. and {Cole}, S. and {Deacon}, N. and {Frenk}, C. and {Fitzsimmons}, A. and {Gezari}, S. and {Gibbs}, V. and {Goessl}, C. and {Goggia}, T. and {Gourgue}, R. and {Goldman}, B. and {Grant}, P. and {Grebel}, E.~K. and {Hambly}, N.~C. and {Hasinger}, G. and {Heavens}, A.~F. and {Heckman}, T.~M. and {Henderson}, R. and {Henning}, T. and {Holman}, M. and {Hopp}, U. and {Ip}, W.-H. and {Isani}, S. and {Jackson}, M. and {Keyes}, C.~D. and {Koekemoer}, A.~M. and {Kotak}, R. and {Le}, D. and {Liska}, D. and {Long}, K.~S. and {Lucey}, J.~R. and {Liu}, M. and {Martin}, N.~F. and {Masci}, G. and {McLean}, B. and {Mindel}, E. and {Misra}, P. and {Morganson}, E. and {Murphy}, D.~N.~A. and {Obaika}, A. and {Narayan}, G. and {Nieto-Santisteban}, M.~A. and {Norberg}, P. and {Peacock}, J.~A. and {Pier}, E.~A. and {Postman}, M. and {Primak}, N. and {Rae}, C. and {Rai}, A. and {Riess}, A. and {Riffeser}, A. and {Rix}, H.~W. and {R{\"o}ser}, S. and {Russel}, R. and {Rutz}, L. and {Schilbach}, E. and {Schultz}, A.~S.~B. and {Scolnic}, D. and {Strolger}, L. and {Szalay}, A. and {Seitz}, S. and {Small}, E. and {Smith}, K.~W. and {Soderblom}, D.~R. and {Taylor}, P. and {Thomson}, R. and {Taylor}, A.~N. and {Thakar}, A.~R. and {Thiel}, J. and {Thilker}, D. and {Unger}, D. and {Urata}, Y. and {Valenti}, J. and {Wagner}, J. and {Walder}, T. and {Walter}, F. and {Watters}, S.~P. and {Werner}, S. and {Wood-Vasey}, W.~M. and {Wyse}, R.},
        title = "{The Pan-STARRS1 Surveys}",
      journal = {arXiv e-prints},
         year = 2016,
        month = dec,
          eid = {arXiv:1612.05560},
        pages = {arXiv:1612.05560},
          doi = {10.48550/arXiv.1612.05560},
archivePrefix = {arXiv},
       eprint = {1612.05560},
 primaryClass = {astro-ph.IM},
       adsurl = {https://ui.adsabs.harvard.edu/abs/2016arXiv161205560C}
}

@ARTICLE{2005ApJ...619L...1M,
       author = {{Martin}, D. Christopher and {Fanson}, James and {Schiminovich}, David and {Morrissey}, Patrick and {Friedman}, Peter G. and {Barlow}, Tom A. and {Conrow}, Tim and {Grange}, Robert and {Jelinsky}, Patrick N. and {Milliard}, Bruno and {Siegmund}, Oswald H.~W. and {Bianchi}, Luciana and {Byun}, Yong-Ik and {Donas}, Jose and {Forster}, Karl and {Heckman}, Timothy M. and {Lee}, Young-Wook and {Madore}, Barry F. and {Malina}, Roger F. and {Neff}, Susan G. and {Rich}, R. Michael and {Small}, Todd and {Surber}, Frank and {Szalay}, Alex S. and {Welsh}, Barry and {Wyder}, Ted K.},
        title = "{The Galaxy Evolution Explorer: A Space Ultraviolet Survey Mission}",
      journal = {\apjl},
         year = 2005,
        month = jan,
       volume = {619},
       number = {1},
        pages = {L1-L6},
          doi = {10.1086/426387},
archivePrefix = {arXiv},
       eprint = {astro-ph/0411302},
 primaryClass = {astro-ph},
       adsurl = {https://ui.adsabs.harvard.edu/abs/2005ApJ...619L...1M}
}

@ARTICLE{2010AJ....140.1868W,
       author = {{Wright}, Edward L. and {Eisenhardt}, Peter R.~M. and {Mainzer}, Amy K. and {Ressler}, Michael E. and {Cutri}, Roc M. and {Jarrett}, Thomas and {Kirkpatrick}, J. Davy and {Padgett}, Deborah and {McMillan}, Robert S. and {Skrutskie}, Michael and {Stanford}, S.~A. and {Cohen}, Martin and {Walker}, Russell G. and {Mather}, John C. and {Leisawitz}, David and {Gautier}, III, Thomas N. and {McLean}, Ian and {Benford}, Dominic and {Lonsdale}, Carol J. and {Blain}, Andrew and {Mendez}, Bryan and {Irace}, William R. and {Duval}, Valerie and {Liu}, Fengchuan and {Royer}, Don and {Heinrichsen}, Ingolf and {Howard}, Joan and {Shannon}, Mark and {Kendall}, Martha and {Walsh}, Amy L. and {Larsen}, Mark and {Cardon}, Joel G. and {Schick}, Scott and {Schwalm}, Mark and {Abid}, Mohamed and {Fabinsky}, Beth and {Naes}, Larry and {Tsai}, Chao-Wei},
        title = "{The Wide-field Infrared Survey Explorer (WISE): Mission Description and Initial On-orbit Performance}",
      journal = {\aj},
         year = 2010,
        month = dec,
       volume = {140},
       number = {6},
        pages = {1868-1881},
          doi = {10.1088/0004-6256/140/6/1868},
archivePrefix = {arXiv},
       eprint = {1008.0031},
 primaryClass = {astro-ph.IM},
       adsurl = {https://ui.adsabs.harvard.edu/abs/2010AJ....140.1868W}
}

@ARTICLE{2017MNRAS.465.3913B,
       author = {{Ben{\'\i}tez-Llambay}, Alejandro and {Navarro}, Julio F. and {Frenk}, Carlos S. and {Sawala}, Till and {Oman}, Kyle and {Fattahi}, Azadeh and {Schaller}, Matthieu and {Schaye}, Joop and {Crain}, Robert A. and {Theuns}, Tom},
        title = "{The properties of `dark' {\ensuremath{\Lambda}}CDM haloes in the Local Group}",
      journal = {\mnras},
         year = 2017,
        month = mar,
       volume = {465},
       number = {4},
        pages = {3913-3926},
          doi = {10.1093/mnras/stw2982},
archivePrefix = {arXiv},
       eprint = {1609.01301},
 primaryClass = {astro-ph.GA},
       adsurl = {https://ui.adsabs.harvard.edu/abs/2017MNRAS.465.3913B}
}

@ARTICLE{2023MNRAS.519.1098C,
       author = {{Catinella}, Barbara and {Cortese}, Luca and {Tiley}, Alfred L. and {Janowiecki}, Steven and {Watts}, Adam B. and {Bryant}, Julia J. and {Croom}, Scott M. and {d'Eugenio}, Francesco and {van de Sande}, Jesse and {Bland-Hawthorn}, Joss and {Fraser-McKelvie}, Amelia and {Richards}, Samuel N. and {Sweet}, Sarah M. and {Pisano}, Daniel J. and {Pingel}, Nickolas and {Koopmann}, Rebecca A. and {Cottrill}, Dillion and {Hill}, Meghan},
        title = "{SAMI-H I: The H I view of the H{\ensuremath{\alpha}} Tully-Fisher relation and data release}",
      journal = {\mnras},
         year = 2023,
        month = feb,
       volume = {519},
       number = {1},
        pages = {1098-1114},
          doi = {10.1093/mnras/stac3556},
archivePrefix = {arXiv},
       eprint = {2212.04728},
 primaryClass = {astro-ph.GA},
       adsurl = {https://ui.adsabs.harvard.edu/abs/2023MNRAS.519.1098C}
}

@ARTICLE{1978MNRAS.183..341W,
       author = {{White}, S.~D.~M. and {Rees}, M.~J.},
        title = "{Core condensation in heavy halos: a two-stage theory for galaxy formation and clustering.}",
      journal = {\mnras},
         year = 1978,
        month = may,
       volume = {183},
        pages = {341-358},
          doi = {10.1093/mnras/183.3.341},
       adsurl = {https://ui.adsabs.harvard.edu/abs/1978MNRAS.183..341W}
}

@ARTICLE{2006MNRAS.371..401H,
       author = {{Hoeft}, Matthias and {Yepes}, Gustavo and {Gottl{\"o}ber}, Stefan and {Springel}, Volker},
        title = "{Dwarf galaxies in voids: suppressing star formation with photoheating}",
      journal = {\mnras},
         year = 2006,
        month = sep,
       volume = {371},
       number = {1},
        pages = {401-414},
          doi = {10.1111/j.1365-2966.2006.10678.x},
archivePrefix = {arXiv},
       eprint = {astro-ph/0501304},
 primaryClass = {astro-ph},
       adsurl = {https://ui.adsabs.harvard.edu/abs/2006MNRAS.371..401H}
}

@ARTICLE{2008MNRAS.390..920O,
       author = {{Okamoto}, Takashi and {Gao}, Liang and {Theuns}, Tom},
        title = "{Mass loss of galaxies due to an ultraviolet background}",
      journal = {\mnras},
         year = 2008,
        month = nov,
       volume = {390},
       number = {3},
        pages = {920-928},
          doi = {10.1111/j.1365-2966.2008.13830.x},
archivePrefix = {arXiv},
       eprint = {0806.0378},
 primaryClass = {astro-ph},
       adsurl = {https://ui.adsabs.harvard.edu/abs/2008MNRAS.390..920O}
}

@ARTICLE{2020MNRAS.498.4887B,
       author = {{Ben{\'\i}tez-Llambay}, Alejandro and {Frenk}, Carlos},
        title = "{The detailed structure and the onset of galaxy formation in low-mass gaseous dark matter haloes}",
      journal = {\mnras},
         year = 2020,
        month = nov,
       volume = {498},
       number = {4},
        pages = {4887-4900},
          doi = {10.1093/mnras/staa2698},
archivePrefix = {arXiv},
       eprint = {2004.06124},
 primaryClass = {astro-ph.GA},
       adsurl = {https://ui.adsabs.harvard.edu/abs/2020MNRAS.498.4887B}
}

@ARTICLE{1992MNRAS.256P..43E,
       author = {{Efstathiou}, G.},
        title = "{Suppressing the formation of dwarf galaxies via photoionization}",
      journal = {\mnras},
         year = 1992,
        month = may,
       volume = {256},
       number = {2},
        pages = {43P-47P},
          doi = {10.1093/mnras/256.1.43P},
       adsurl = {https://ui.adsabs.harvard.edu/abs/1992MNRAS.256P..43E}
}

@ARTICLE{2016MNRAS.457.1931S,
       author = {{Sawala}, Till and {Frenk}, Carlos S. and {Fattahi}, Azadeh and {Navarro}, Julio F. and {Bower}, Richard G. and {Crain}, Robert A. and {Dalla Vecchia}, Claudio and {Furlong}, Michelle and {Helly}, John. C. and {Jenkins}, Adrian and {Oman}, Kyle A. and {Schaller}, Matthieu and {Schaye}, Joop and {Theuns}, Tom and {Trayford}, James and {White}, Simon D.~M.},
        title = "{The APOSTLE simulations: solutions to the Local Group's cosmic puzzles}",
      journal = {\mnras},
         year = 2016,
        month = apr,
       volume = {457},
       number = {2},
        pages = {1931-1943},
          doi = {10.1093/mnras/stw145},
archivePrefix = {arXiv},
       eprint = {1511.01098},
 primaryClass = {astro-ph.GA},
       adsurl = {https://ui.adsabs.harvard.edu/abs/2016MNRAS.457.1931S}
}

@ARTICLE{2002MNRAS.336..541V,
       author = {{Verde}, Licia and {Oh}, S. Peng and {Jimenez}, Raul},
        title = "{The abundance of dark galaxies}",
      journal = {\mnras},
         year = 2002,
        month = oct,
       volume = {336},
       number = {2},
        pages = {541-549},
          doi = {10.1046/j.1365-8711.2002.05771.x},
archivePrefix = {arXiv},
       eprint = {astro-ph/0202283},
 primaryClass = {astro-ph},
       adsurl = {https://ui.adsabs.harvard.edu/abs/2002MNRAS.336..541V}
}

@ARTICLE{2024ApJ...962..129L,
       author = {{Lee}, Gain and {Hwang}, Ho Seong and {Lee}, Jaehyun and {Shin}, Jihye and {Song}, Hyunmi},
        title = "{Understanding the Formation and Evolution of Dark Galaxies in a Simulated Universe}",
      journal = {\apj},
         year = 2024,
        month = feb,
       volume = {962},
       number = {2},
          eid = {129},
        pages = {129},
          doi = {10.3847/1538-4357/ad1e5d},
archivePrefix = {arXiv},
       eprint = {2401.07007},
 primaryClass = {astro-ph.GA},
       adsurl = {https://ui.adsabs.harvard.edu/abs/2024ApJ...962..129L}
}

@ARTICLE{2026arXiv260427047M,
       author = {{Moreno}, Jorge and {Wheeler}, Coral and {Mercado}, Francisco J. and {Rodriguez Wimberly}, M. Katy and {Gandhi}, Pratik J. and {Samuel}, Jenna and {Feldmann}, Robert and {Bullock}, James S. and {Wetzel}, Andrew and {Boylan-Kolchin}, Michael and {Hopkins}, Philip F.},
        title = "{Beyond Cloud-9: The case for discovering more HI-rich failed halos}",
      journal = {arXiv e-prints},
         year = 2026,
        month = apr,
          eid = {arXiv:2604.27047},
        pages = {arXiv:2604.27047},
          doi = {10.48550/arXiv.2604.27047},
archivePrefix = {arXiv},
       eprint = {2604.27047},
 primaryClass = {astro-ph.GA},
       adsurl = {https://ui.adsabs.harvard.edu/abs/2026arXiv260427047M}
}

@ARTICLE{2020MNRAS.498.2968L,
       author = {{Libeskind}, Noam I. and {Carlesi}, Edoardo and {Grand}, Robert J.~J. and {Khalatyan}, Arman and {Knebe}, Alexander and {Pakmor}, Ruediger and {Pilipenko}, Sergey and {Pawlowski}, Marcel S. and {Sparre}, Martin and {Tempel}, Elmo and {Wang}, Peng and {Courtois}, H{\'e}l{\`e}ne M. and {Gottl{\"o}ber}, Stefan and {Hoffman}, Yehuda and {Minchev}, Ivan and {Pfrommer}, Christoph and {Sorce}, Jenny G. and {Springel}, Volker and {Steinmetz}, Matthias and {Tully}, R. Brent and {Vogelsberger}, Mark and {Yepes}, Gustavo},
        title = "{The HESTIA project: simulations of the Local Group}",
      journal = {\mnras},
         year = 2020,
        month = oct,
       volume = {498},
       number = {2},
        pages = {2968-2983},
          doi = {10.1093/mnras/staa2541},
archivePrefix = {arXiv},
       eprint = {2008.04926},
 primaryClass = {astro-ph.GA},
       adsurl = {https://ui.adsabs.harvard.edu/abs/2020MNRAS.498.2968L}
}

@ARTICLE{2017MNRAS.467..179G,
       author = {{Grand}, Robert J.~J. and {G{\'o}mez}, Facundo A. and {Marinacci}, Federico and {Pakmor}, R{\"u}diger and {Springel}, Volker and {Campbell}, David J.~R. and {Frenk}, Carlos S. and {Jenkins}, Adrian and {White}, Simon D.~M.},
        title = "{The Auriga Project: the properties and formation mechanisms of disc galaxies across cosmic time}",
      journal = {\mnras},
         year = 2017,
        month = may,
       volume = {467},
       number = {1},
        pages = {179-207},
          doi = {10.1093/mnras/stx071},
archivePrefix = {arXiv},
       eprint = {1610.01159},
 primaryClass = {astro-ph.GA},
       adsurl = {https://ui.adsabs.harvard.edu/abs/2017MNRAS.467..179G}
}

@ARTICLE{2023ApJ...944L..40X,
       author = {{Xu}, Jin-Long and {Zhu}, Ming and {Yu}, Naiping and {Zhang}, Chuan-Peng and {Liu}, Xiao-Lan and {Ai}, Mei and {Jiang}, Peng},
        title = "{Discovery of an Isolated Dark Dwarf Galaxy in the Nearby Universe}",
      journal = {\apjl},
         year = 2023,
        month = feb,
       volume = {944},
       number = {2},
          eid = {L40},
        pages = {L40},
          doi = {10.3847/2041-8213/acb932},
archivePrefix = {arXiv},
       eprint = {2302.02646},
 primaryClass = {astro-ph.CO},
       adsurl = {https://ui.adsabs.harvard.edu/abs/2023ApJ...944L..40X}
}

@ARTICLE{1986MNRAS.218P..25R,
       author = {{Rees}, M.~J.},
        title = "{Lyman absorption lines in quasar spectra - Evidence for gravitationally-confined gas in dark minihaloes}",
      journal = {\mnras},
         year = 1986,
        month = jan,
       volume = {218},
        pages = {25P-30P},
          doi = {10.1093/mnras/218.1.25P},
       adsurl = {https://ui.adsabs.harvard.edu/abs/1986MNRAS.218P..25R}
}

@ARTICLE{1986Ap&SS.118..509I,
       author = {{Ikeuchi}, S.},
        title = "{The baryon clump within an extended dark matter region}",
      journal = {\apss},
         year = 1986,
        month = jan,
       volume = {118},
       number = {1-2},
        pages = {509-514},
          doi = {10.1007/BF00651178},
       adsurl = {https://ui.adsabs.harvard.edu/abs/1986Ap&SS.118..509I}
}

@ARTICLE{2026A&A...708A..40S,
       author = {{{\v{S}}iljeg}, Barbara and {Adams}, Elizabeth A.~K. and {Oosterloo}, Tom A. and {Fraternali}, Filippo and {Hess}, Kelley M. and {Xu}, Jin-Long and {Zhu}, Ming},
        title = "{Not so-dark: High resolution H I imaging of J0139+4328 and identification of an optical counterpart}",
      journal = {\aap},
         year = 2026,
        month = mar,
       volume = {708},
          eid = {A40},
        pages = {A40},
          doi = {10.1051/0004-6361/202556900},
archivePrefix = {arXiv},
       eprint = {2601.12513},
 primaryClass = {astro-ph.GA},
       adsurl = {https://ui.adsabs.harvard.edu/abs/2026A&A...708A..40S}
}

@ARTICLE{2019AJ....157..168D,
       author = {{Dey}, Arjun and {Schlegel}, David J. and {Lang}, Dustin and {Blum}, Robert and {Burleigh}, Kaylan and {Fan}, Xiaohui and {Findlay}, Joseph R. and {Finkbeiner}, Doug and {Herrera}, David and {Juneau}, St{\'e}phanie and {Landriau}, Martin and {Levi}, Michael and {McGreer}, Ian and {Meisner}, Aaron and {Myers}, Adam D. and {Moustakas}, John and {Nugent}, Peter and {Patej}, Anna and {Schlafly}, Edward F. and {Walker}, Alistair R. and {Valdes}, Francisco and {Weaver}, Benjamin A. and {Y{\`e}che}, Christophe and {Zou}, Hu and {Zhou}, Xu and {Abareshi}, Behzad and {Abbott}, T.~M.~C. and {Abolfathi}, Bela and {Aguilera}, C. and {Alam}, Shadab and {Allen}, Lori and {Alvarez}, A. and {Annis}, James and {Ansarinejad}, Behzad and {Aubert}, Marie and {Beechert}, Jacqueline and {Bell}, Eric F. and {BenZvi}, Segev Y. and {Beutler}, Florian and {Bielby}, Richard M. and {Bolton}, Adam S. and {Brice{\~n}o}, C{\'e}sar and {Buckley-Geer}, Elizabeth J. and {Butler}, Karen and {Calamida}, Annalisa and {Carlberg}, Raymond G. and {Carter}, Paul and {Casas}, Ricard and {Castander}, Francisco J. and {Choi}, Yumi and {Comparat}, Johan and {Cukanovaite}, Elena and {Delubac}, Timoth{\'e}e and {DeVries}, Kaitlin and {Dey}, Sharmila and {Dhungana}, Govinda and {Dickinson}, Mark and {Ding}, Zhejie and {Donaldson}, John B. and {Duan}, Yutong and {Duckworth}, Christopher J. and {Eftekharzadeh}, Sarah and {Eisenstein}, Daniel J. and {Etourneau}, Thomas and {Fagrelius}, Parker A. and {Farihi}, Jay and {Fitzpatrick}, Mike and {Font-Ribera}, Andreu and {Fulmer}, Leah and {G{\"a}nsicke}, Boris T. and {Gaztanaga}, Enrique and {George}, Koshy and {Gerdes}, David W. and {Gontcho}, Satya Gontcho A. and {Gorgoni}, Claudio and {Green}, Gregory and {Guy}, Julien and {Harmer}, Diane and {Hernandez}, M. and {Honscheid}, Klaus and {Huang}, Lijuan Wendy and {James}, David J. and {Jannuzi}, Buell T. and {Jiang}, Linhua and {Joyce}, Richard and {Karcher}, Armin and {Karkar}, Sonia and {Kehoe}, Robert and {Kneib}, Jean-Paul and {Kueter-Young}, Andrea and {Lan}, Ting-Wen and {Lauer}, Tod R. and {Le Guillou}, Laurent and {Le Van Suu}, Auguste and {Lee}, Jae Hyeon and {Lesser}, Michael and {Perreault Levasseur}, Laurence and {Li}, Ting S. and {Mann}, Justin L. and {Marshall}, Robert and {Mart{\'\i}nez-V{\'a}zquez}, C.~E. and {Martini}, Paul and {du Mas des Bourboux}, H{\'e}lion and {McManus}, Sean and {Meier}, Tobias Gabriel and {M{\'e}nard}, Brice and {Metcalfe}, Nigel and {Mu{\~n}oz-Guti{\'e}rrez}, Andrea and {Najita}, Joan and {Napier}, Kevin and {Narayan}, Gautham and {Newman}, Jeffrey A. and {Nie}, Jundan and {Nord}, Brian and {Norman}, Dara J. and {Olsen}, Knut A.~G. and {Paat}, Anthony and {Palanque-Delabrouille}, Nathalie and {Peng}, Xiyan and {Poppett}, Claire L. and {Poremba}, Megan R. and {Prakash}, Abhishek and {Rabinowitz}, David and {Raichoor}, Anand and {Rezaie}, Mehdi and {Robertson}, A.~N. and {Roe}, Natalie A. and {Ross}, Ashley J. and {Ross}, Nicholas P. and {Rudnick}, Gregory and {Safonova}, Sasha and {Saha}, Abhijit and {S{\'a}nchez}, F. Javier and {Savary}, Elodie and {Schweiker}, Heidi and {Scott}, Adam and {Seo}, Hee-Jong and {Shan}, Huanyuan and {Silva}, David R. and {Slepian}, Zachary and {Soto}, Christian and {Sprayberry}, David and {Staten}, Ryan and {Stillman}, Coley M. and {Stupak}, Robert J. and {Summers}, David L. and {Sien Tie}, Suk and {Tirado}, H. and {Vargas-Maga{\~n}a}, Mariana and {Vivas}, A. Katherina and {Wechsler}, Risa H. and {Williams}, Doug and {Yang}, Jinyi and {Yang}, Qian and {Yapici}, Tolga and {Zaritsky}, Dennis and {Zenteno}, A. and {Zhang}, Kai and {Zhang}, Tianmeng and {Zhou}, Rongpu and {Zhou}, Zhimin},
        title = "{Overview of the DESI Legacy Imaging Surveys}",
      journal = {\aj},
         year = 2019,
        month = may,
       volume = {157},
       number = {5},
          eid = {168},
        pages = {168},
          doi = {10.3847/1538-3881/ab089d},
archivePrefix = {arXiv},
       eprint = {1804.08657},
 primaryClass = {astro-ph.IM},
       adsurl = {https://ui.adsabs.harvard.edu/abs/2019AJ....157..168D}
}


\bsp	
\label{lastpage}
\end{document}


\section*{Mock galaxy injections}

\begin{figure*}
    \centering
    \includegraphics[width=\textwidth]{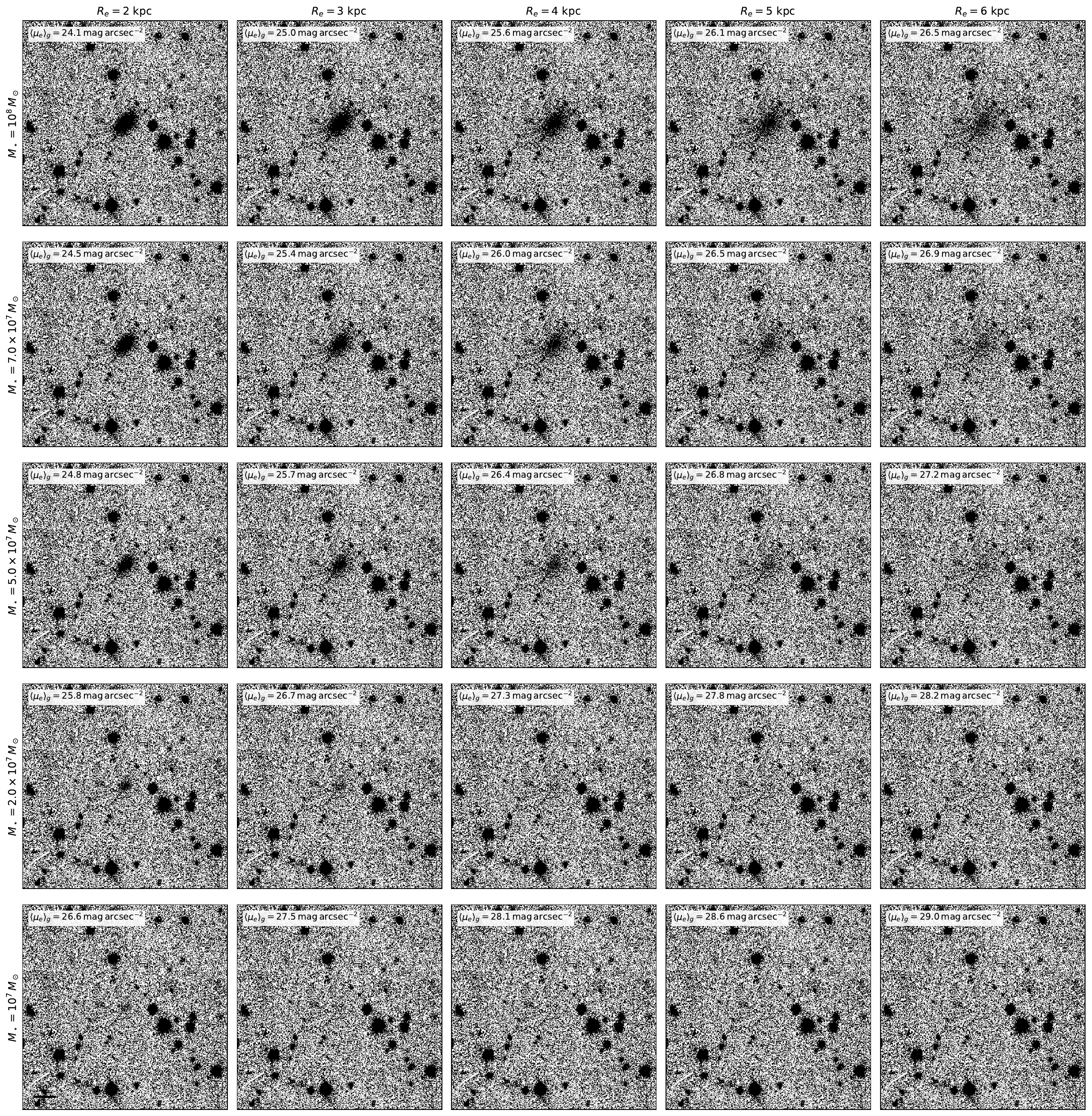}
    \caption{Mock galaxy injection run with $M/L_g = 0.5\ M_\odot/L_\odot$. The injected galaxies are tilted exponential discs (S\'ersic index $= 1$, inclination $i = 60\degr$), assuming an intrinsic axial ratio $q_0 = 0.20$. The effective radius $R_e$ increases along the columns, while the stellar mass $M_\bigstar$ decreases along the rows, so that the surface brightness increases towards the top-left corner. In each panel the mean surface brightness of the galaxy is indicated. In this case we injected the galaxies within the shared region between the radio beams (i.e. the most probable position for the HI source).}
    \label{fig:supp1}
\end{figure*}

\clearpage

\begin{figure*}
    \centering
    \includegraphics[width=\textwidth]{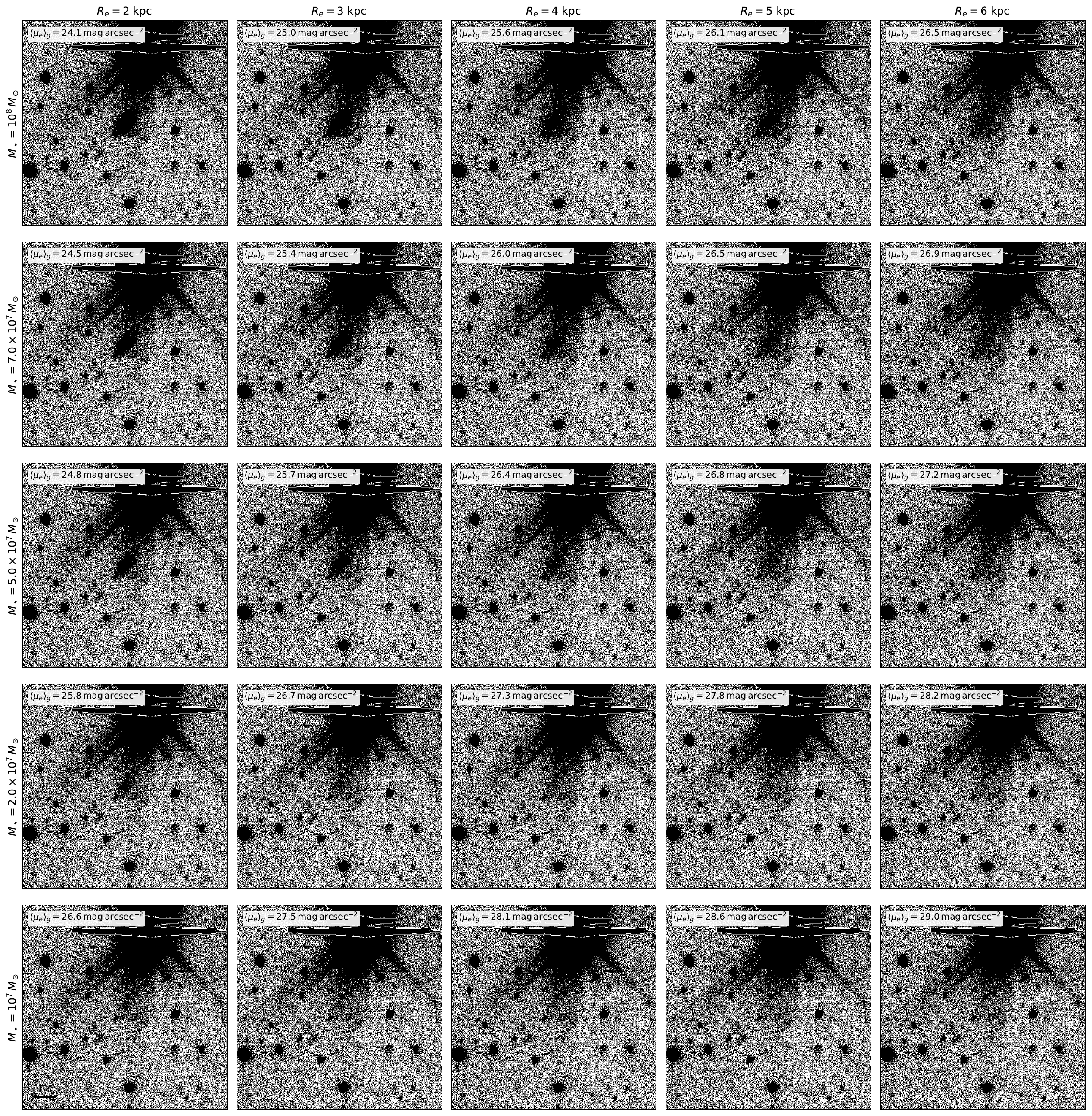}
    \caption{The same as Figure \ref{fig:supp1}, but with the galaxies injected into the glare of the two bright foreground stars. This is a worst-case scenario: this position is incompatible with the FASHI registered position for AGC 322753.}
    \label{fig:supp2}
\end{figure*}

\clearpage

\begin{figure*}
    \centering
    \includegraphics[width=\textwidth]{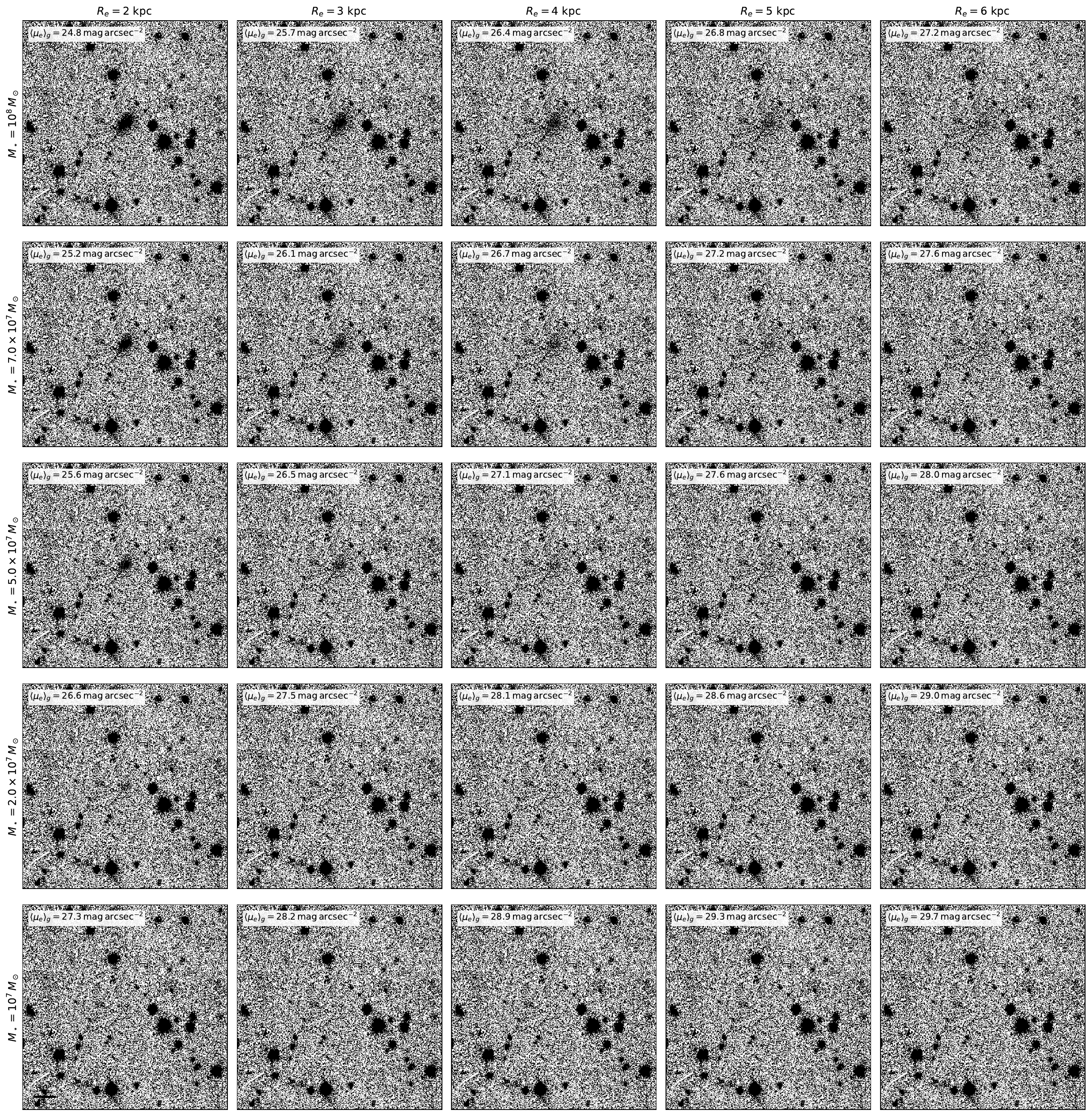}
    \caption{The same as Figure \ref{fig:supp1} but with $M/L_g = 1\ M_\odot/L_\odot$.}
    \label{fig:supp3}
\end{figure*}

\clearpage

\begin{figure*}
    \centering
    \includegraphics[width=\textwidth]{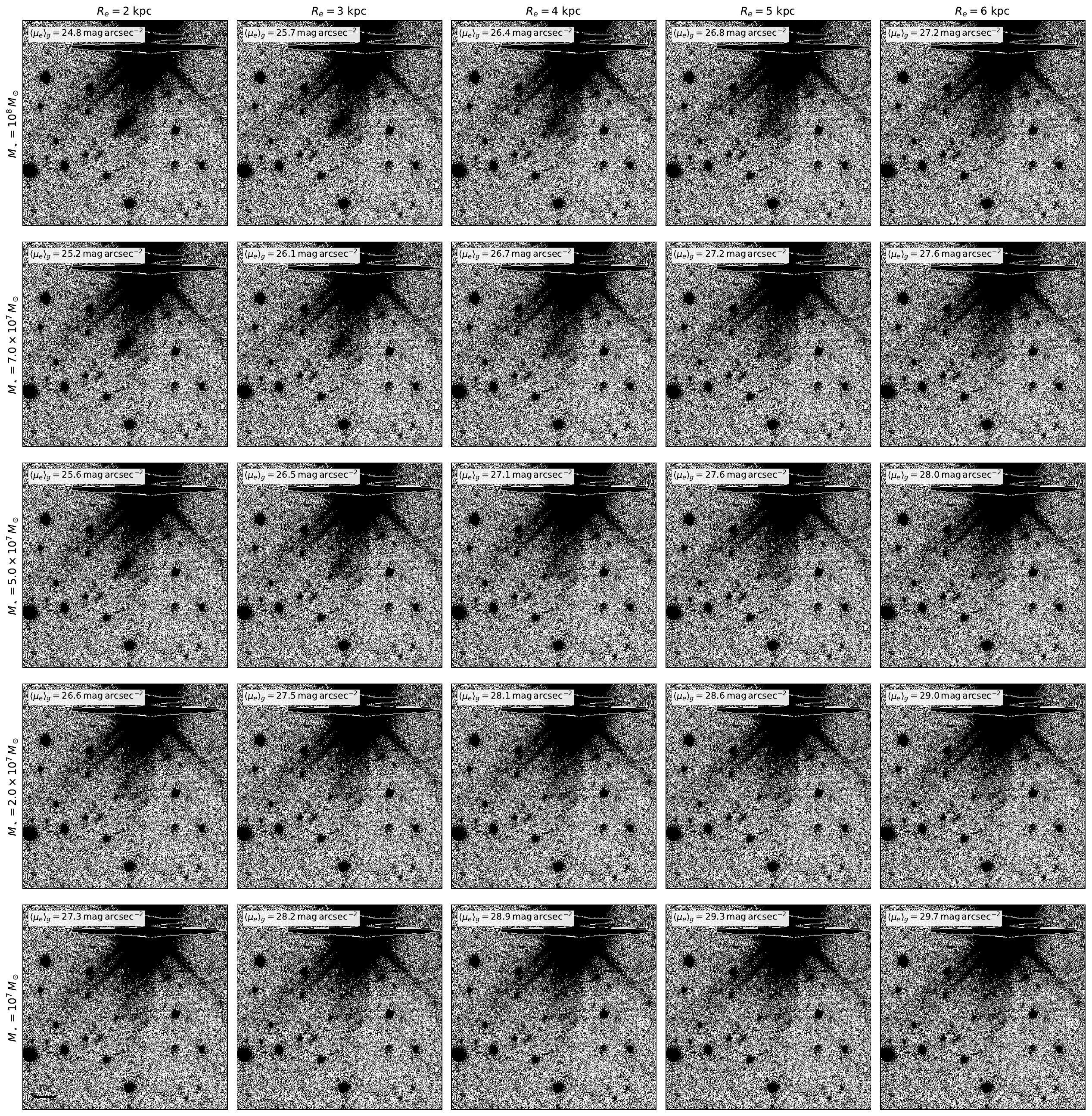}
    \caption{The same as Figure \ref{fig:supp2} but with $M/L_g = 1\ M_\odot/L_\odot$.}
    \label{fig:supp4}
\end{figure*}

\clearpage

\section*{Discarded sources}

\begin{figure*}
    \centering
    \includegraphics[width=\textwidth]{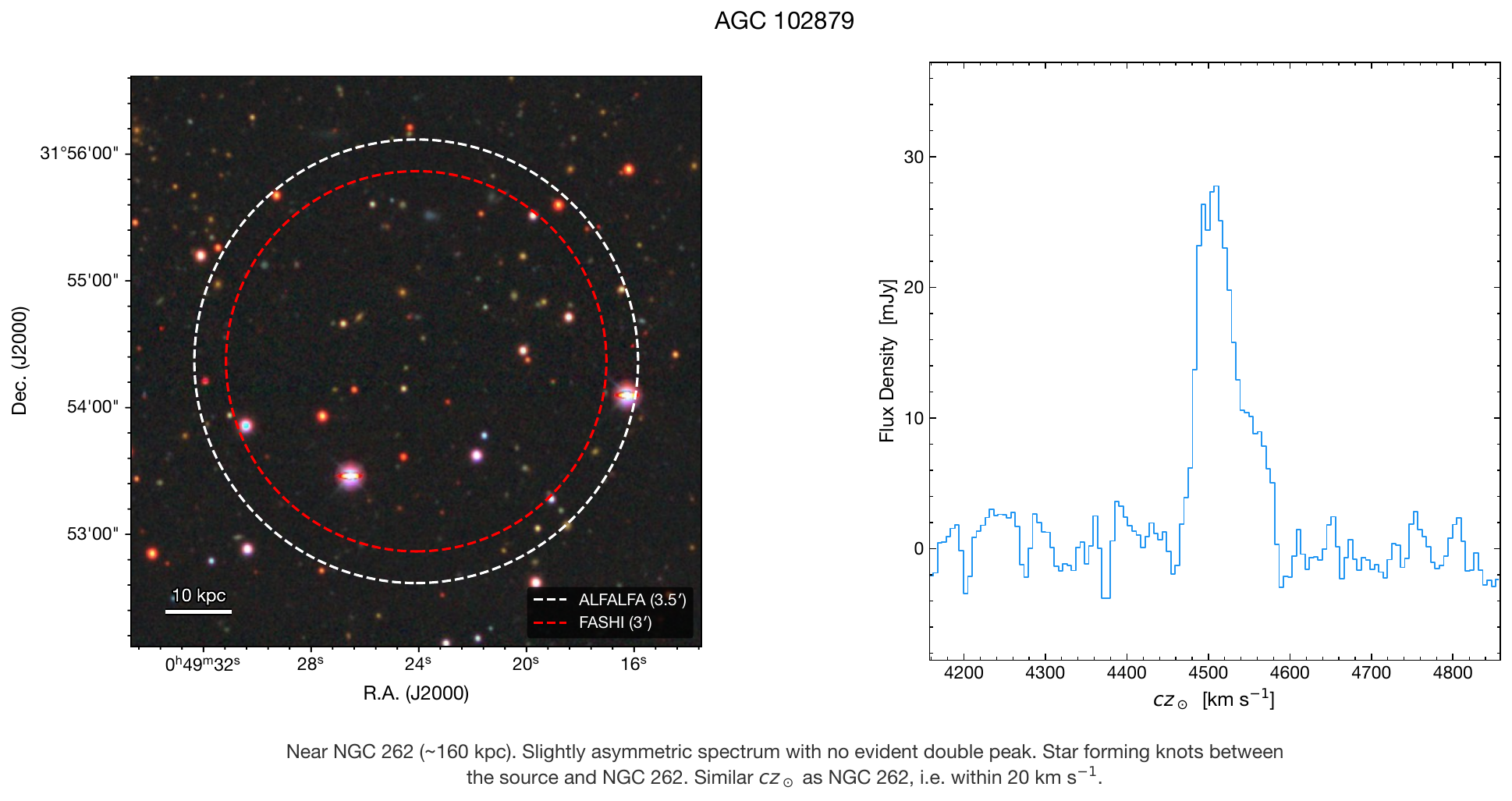}
\end{figure*}

\begin{figure*}
    \centering
    \includegraphics[width=\textwidth]{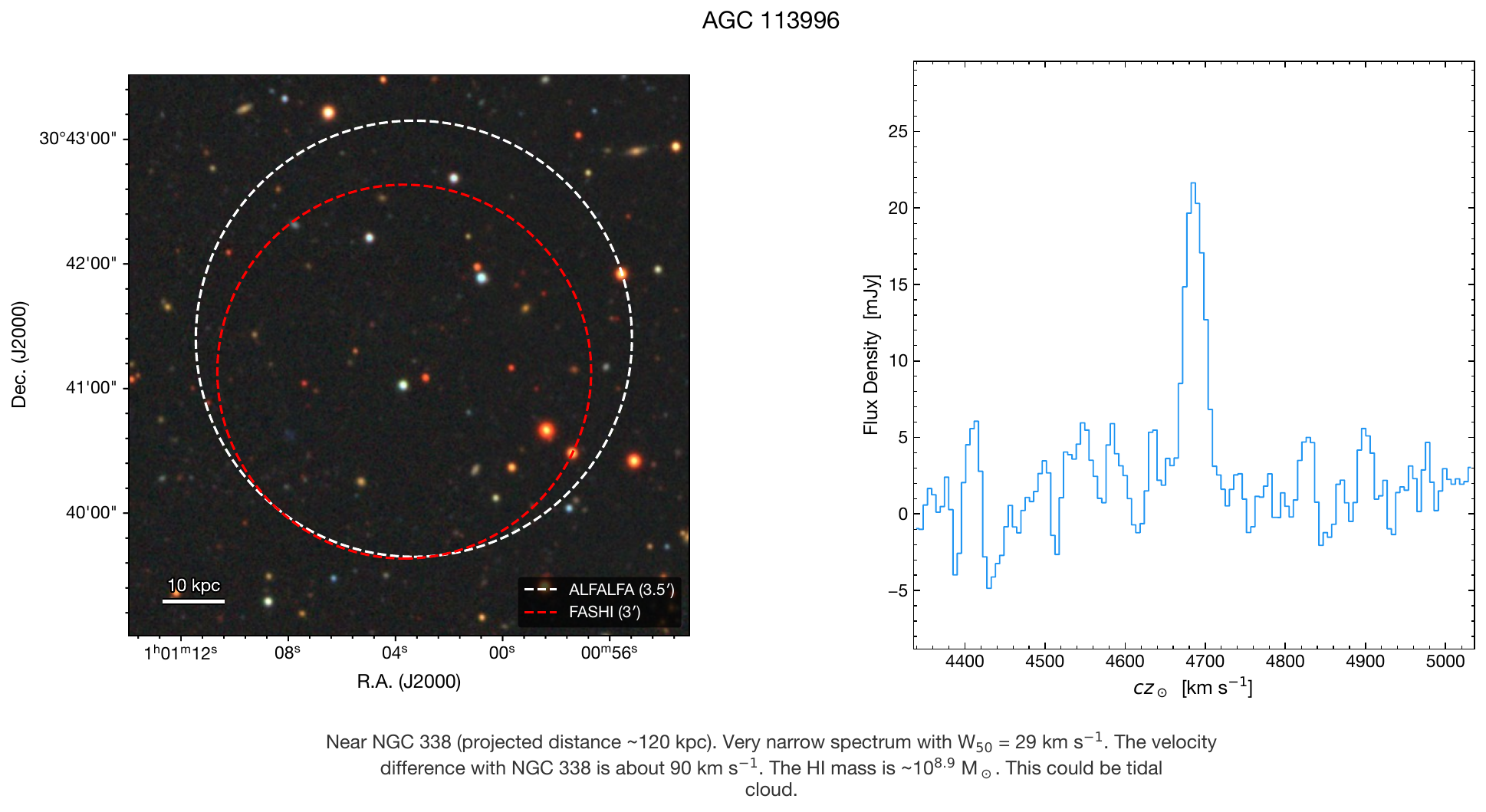}
\end{figure*}

\begin{figure*}
    \centering
    \includegraphics[width=\textwidth]{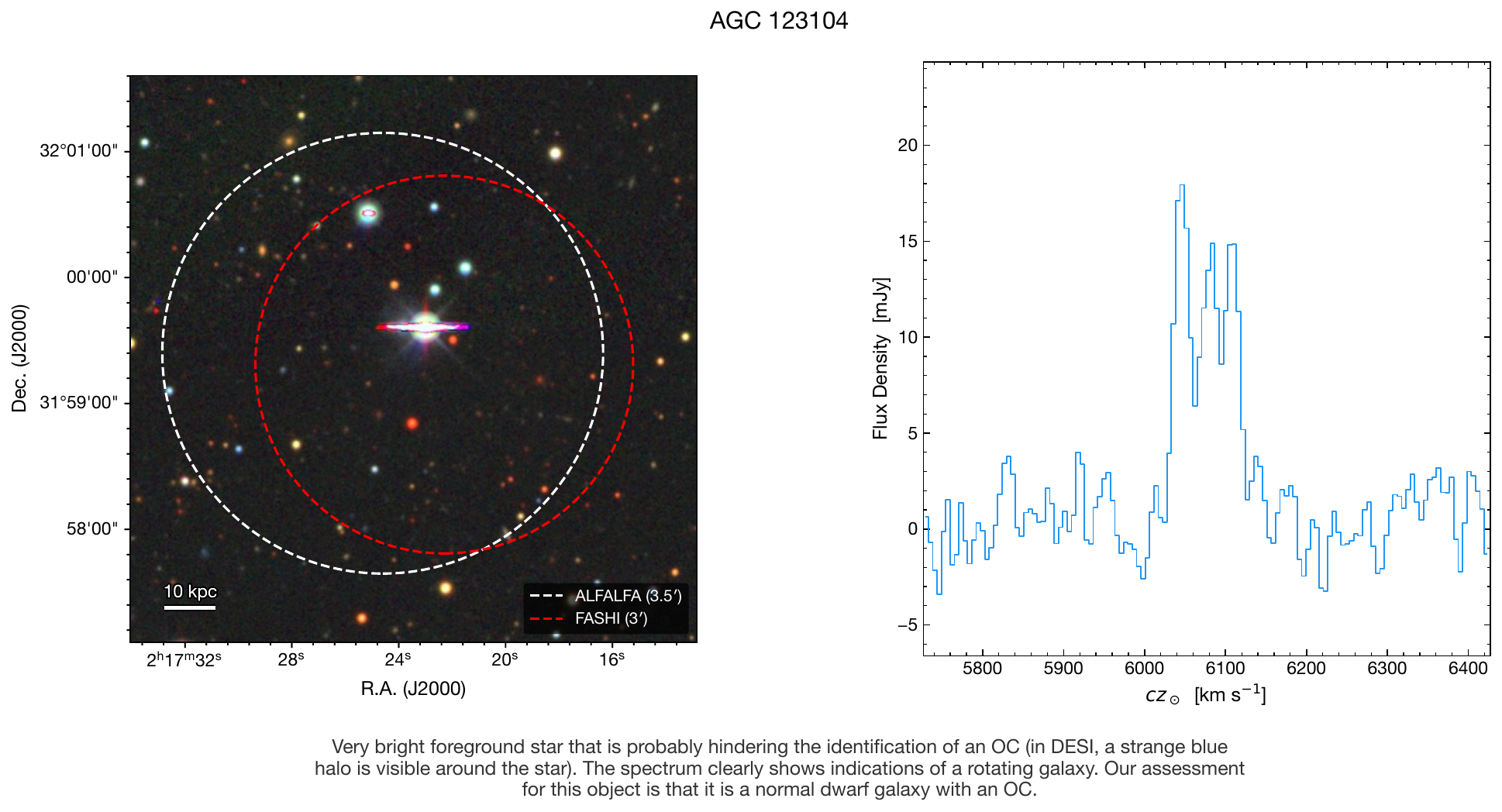}
\end{figure*}

\begin{figure*}
    \centering
    \includegraphics[width=\textwidth]{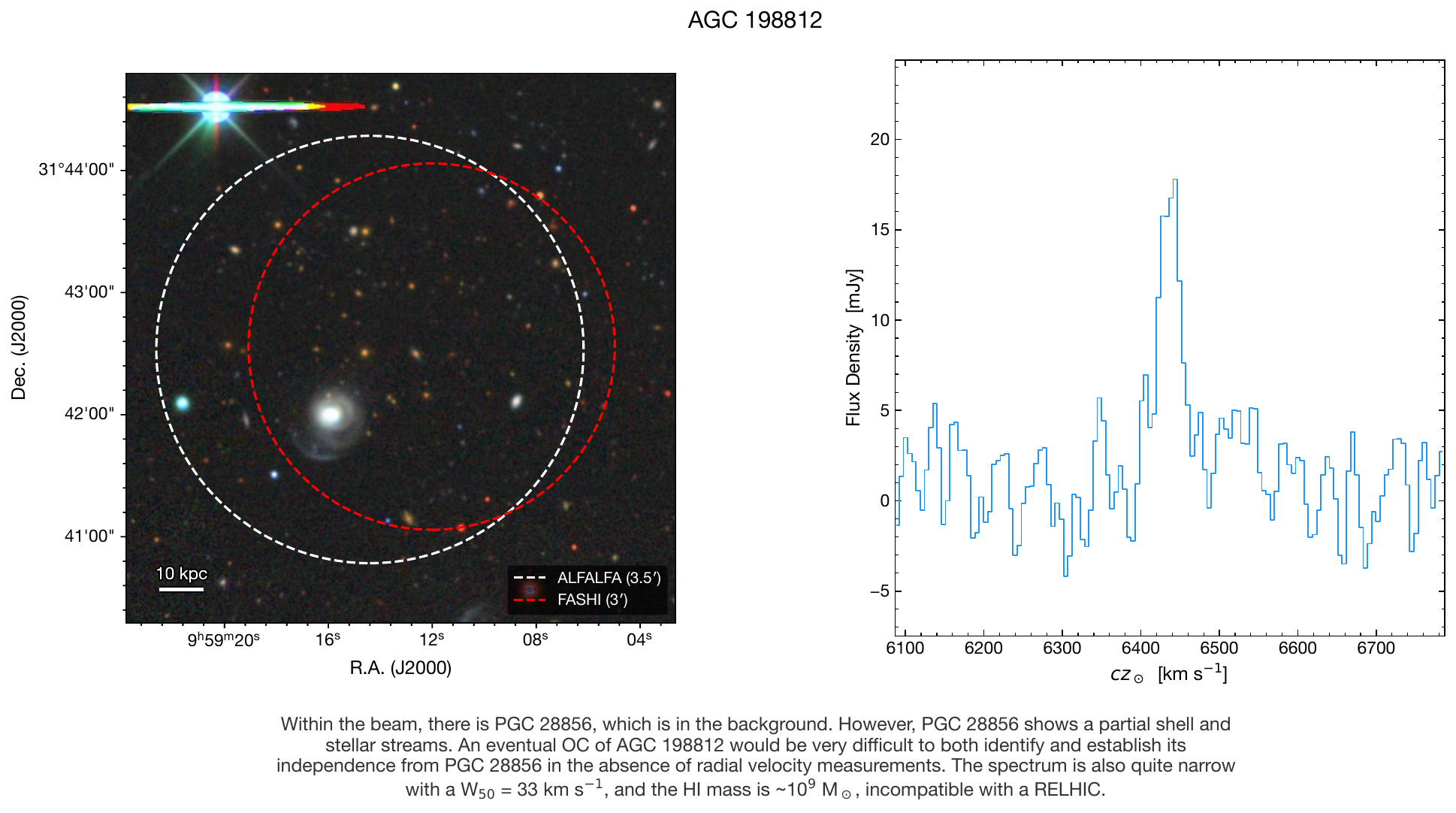}
\end{figure*}

\begin{figure*}
    \centering
    \includegraphics[width=\textwidth]{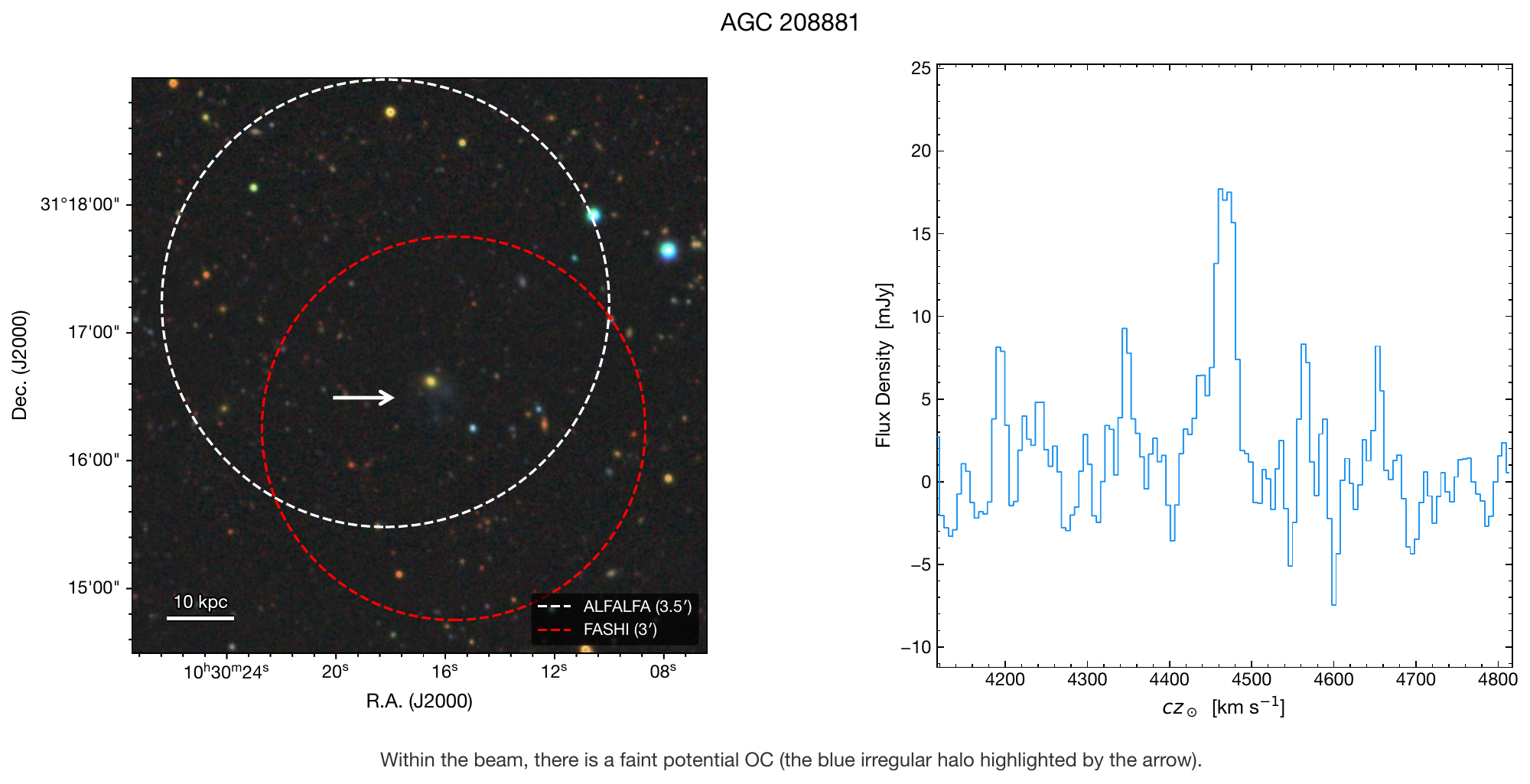}
\end{figure*}

\begin{figure*}
    \centering
    \includegraphics[width=\textwidth]{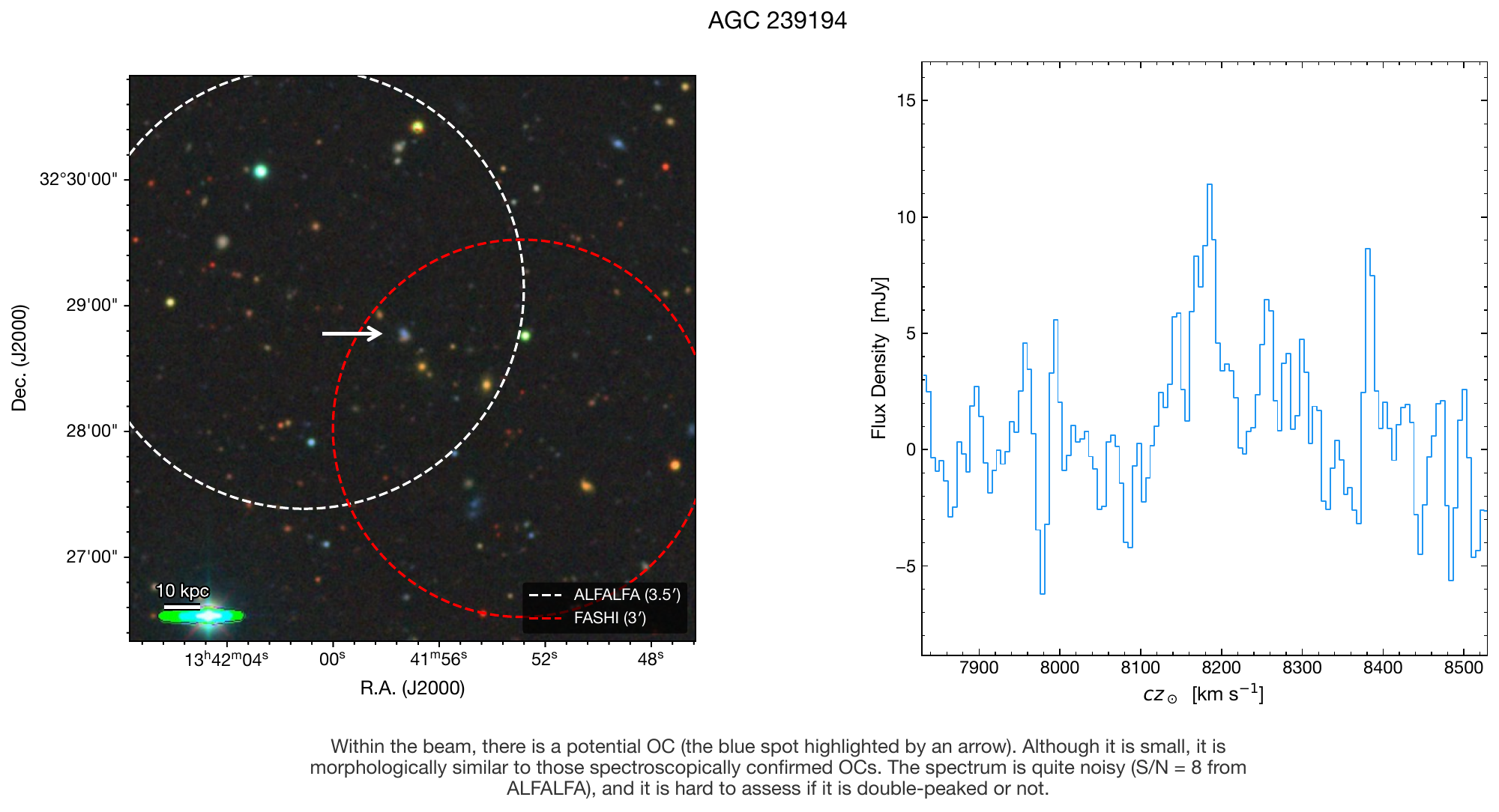}
\end{figure*}

\begin{figure*}
    \centering
    \includegraphics[width=\textwidth]{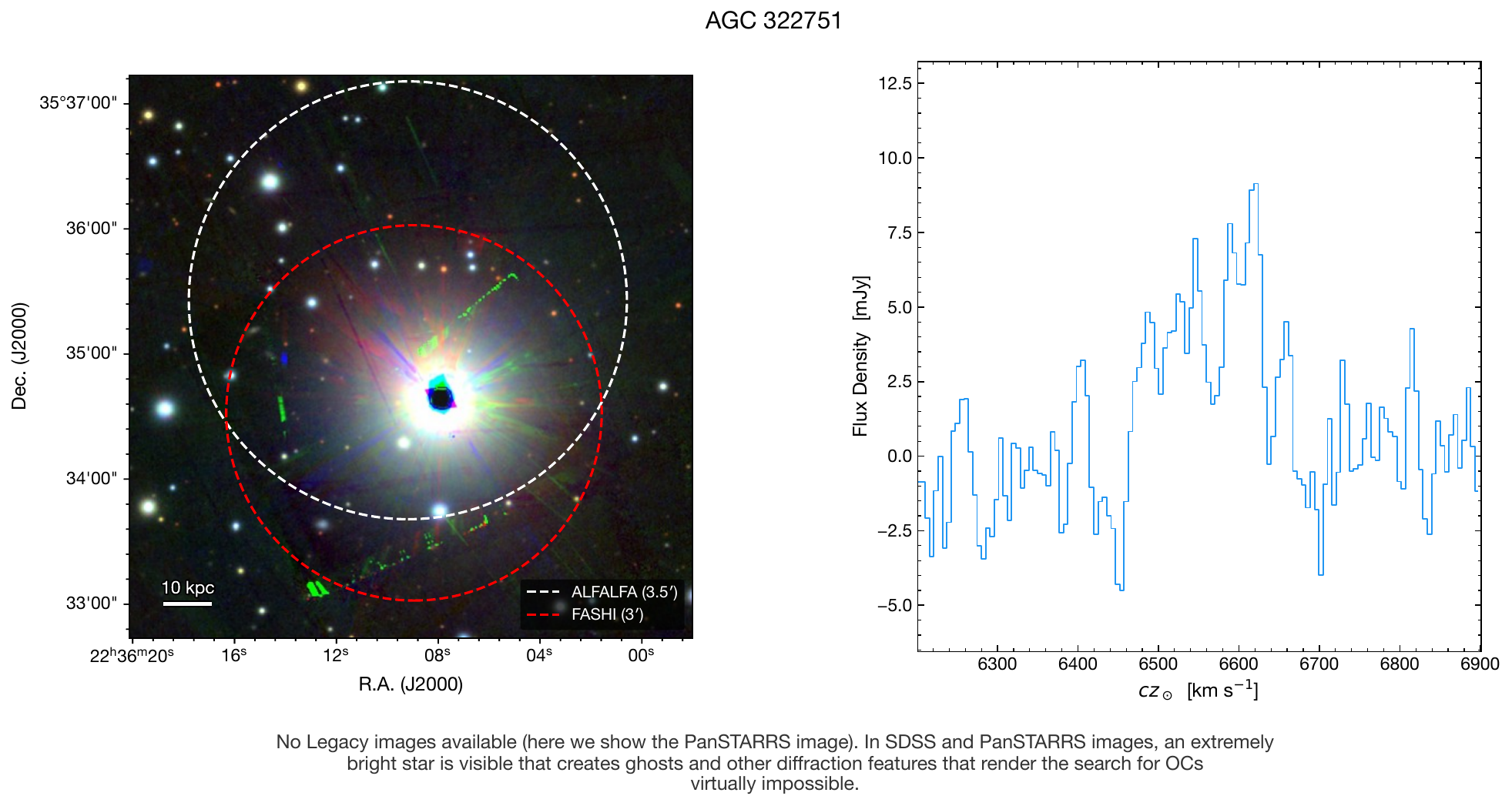}
\end{figure*}

\begin{figure*}
    \centering
    \includegraphics[width=\textwidth]{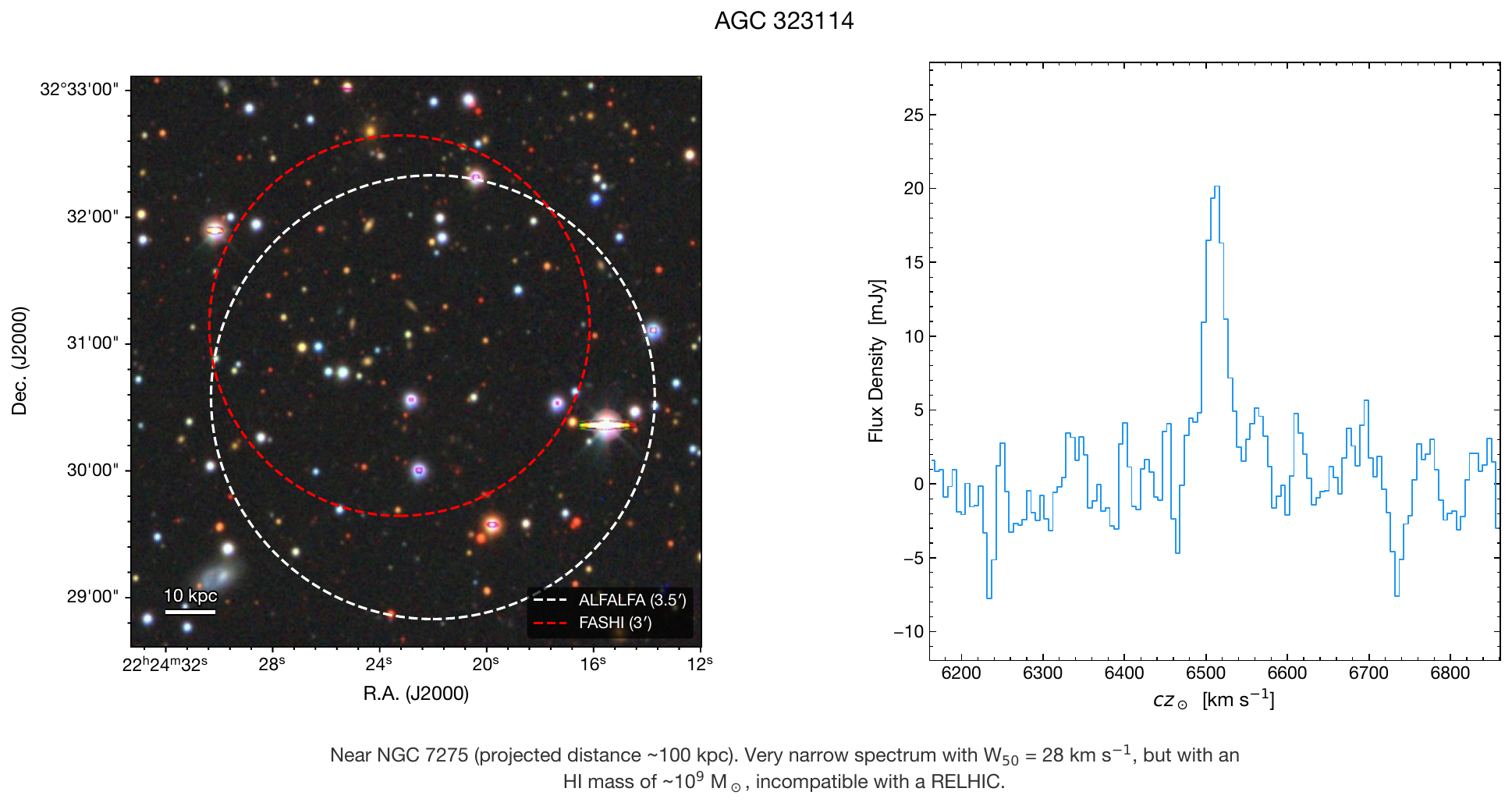}
\end{figure*}

\begin{figure*}
    \centering
    \includegraphics[width=\textwidth]{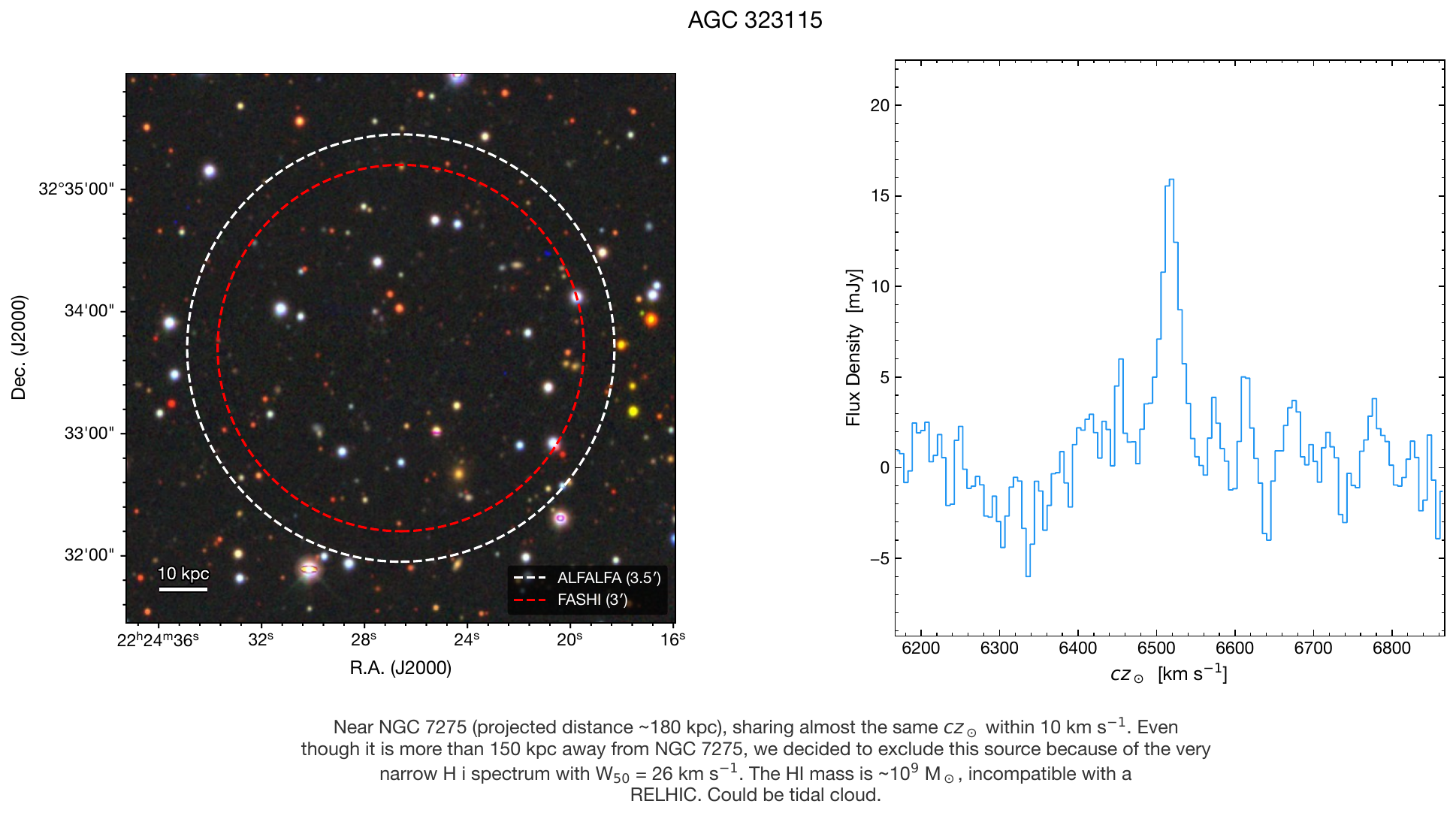}
\end{figure*}

\begin{figure*}
    \centering
    \includegraphics[width=\textwidth]{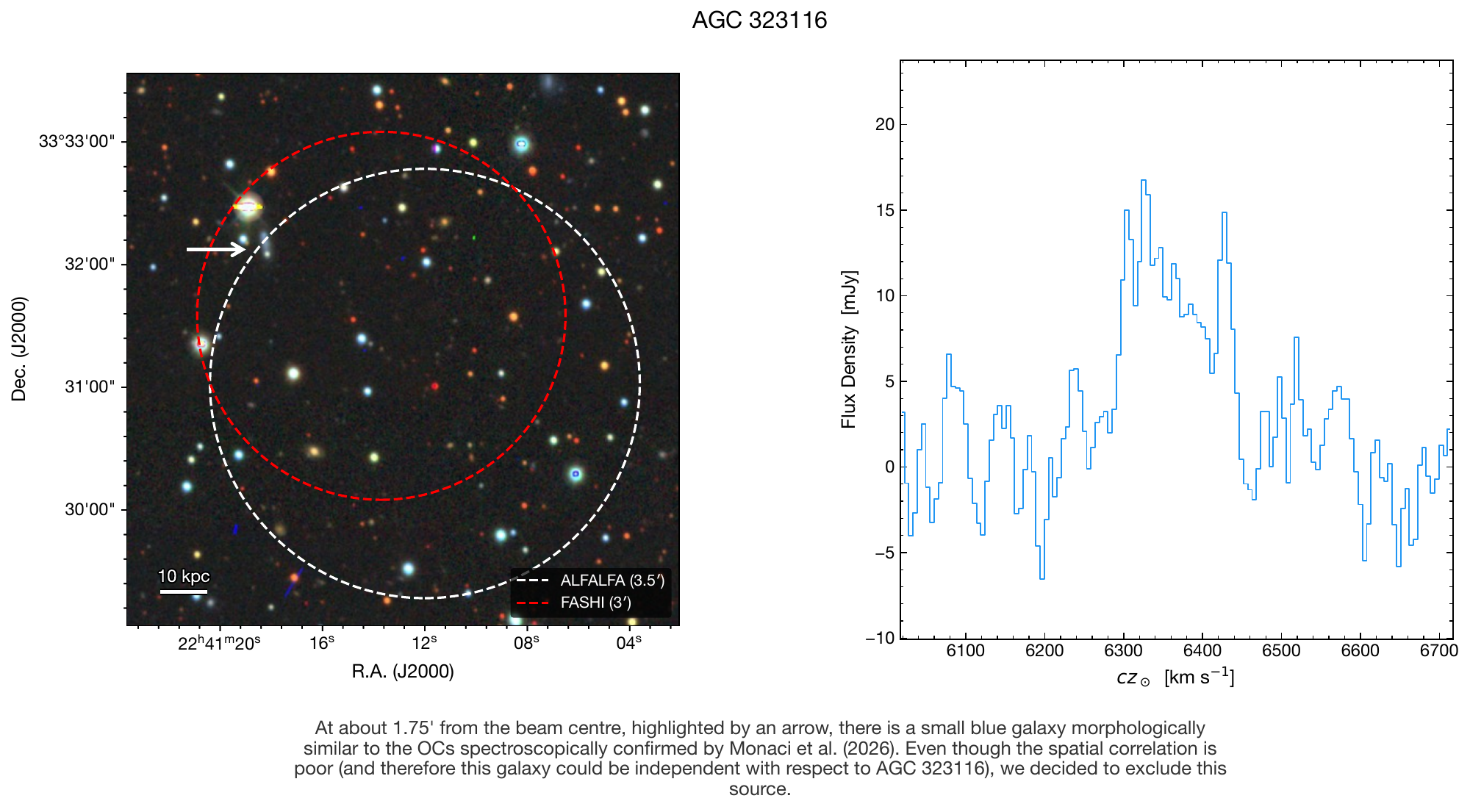}
\end{figure*}

\begin{figure*}
    \centering
    \includegraphics[width=\textwidth]{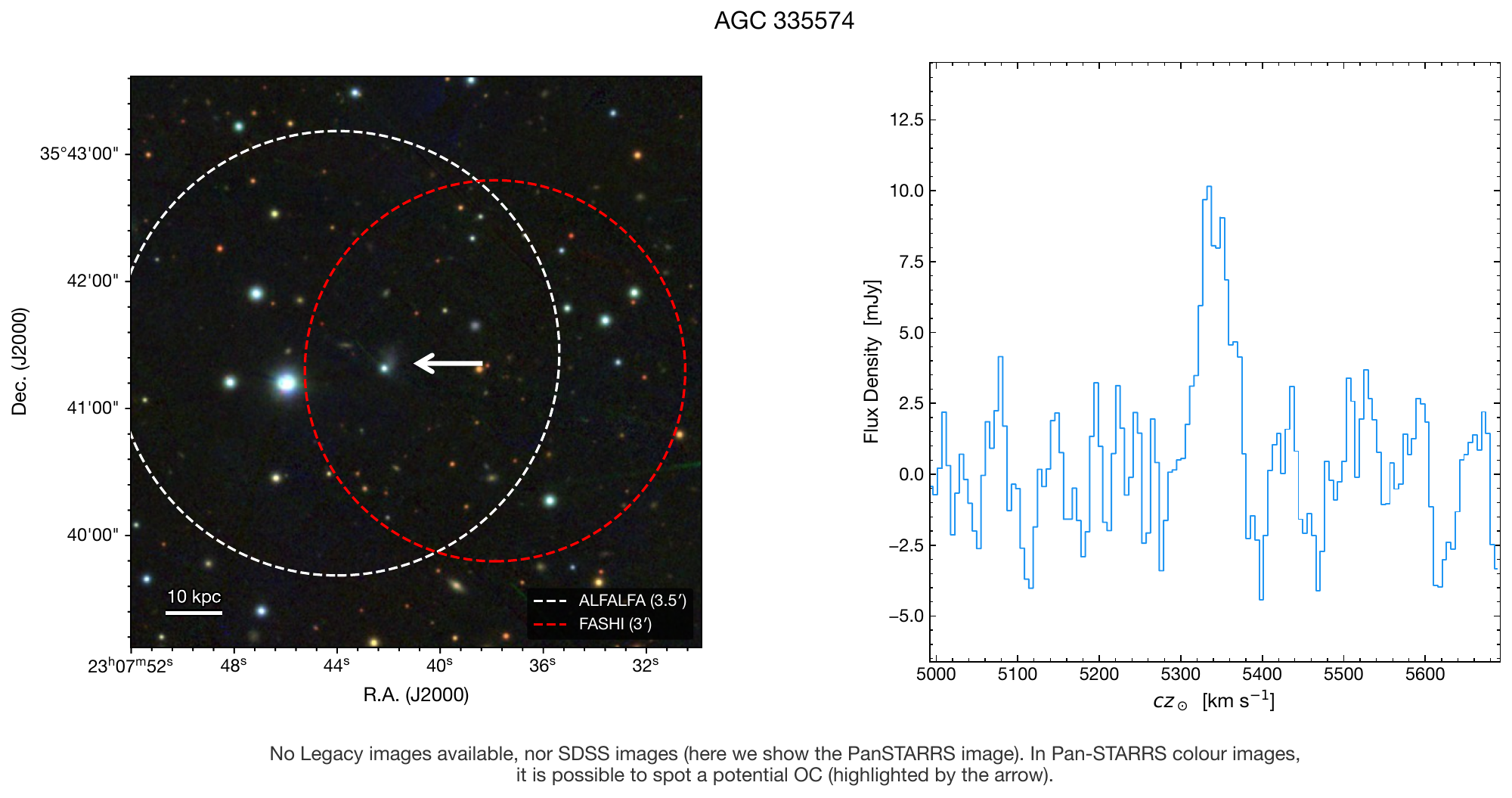}
\end{figure*}

\begin{figure*}
    \centering
    \includegraphics[width=\textwidth]{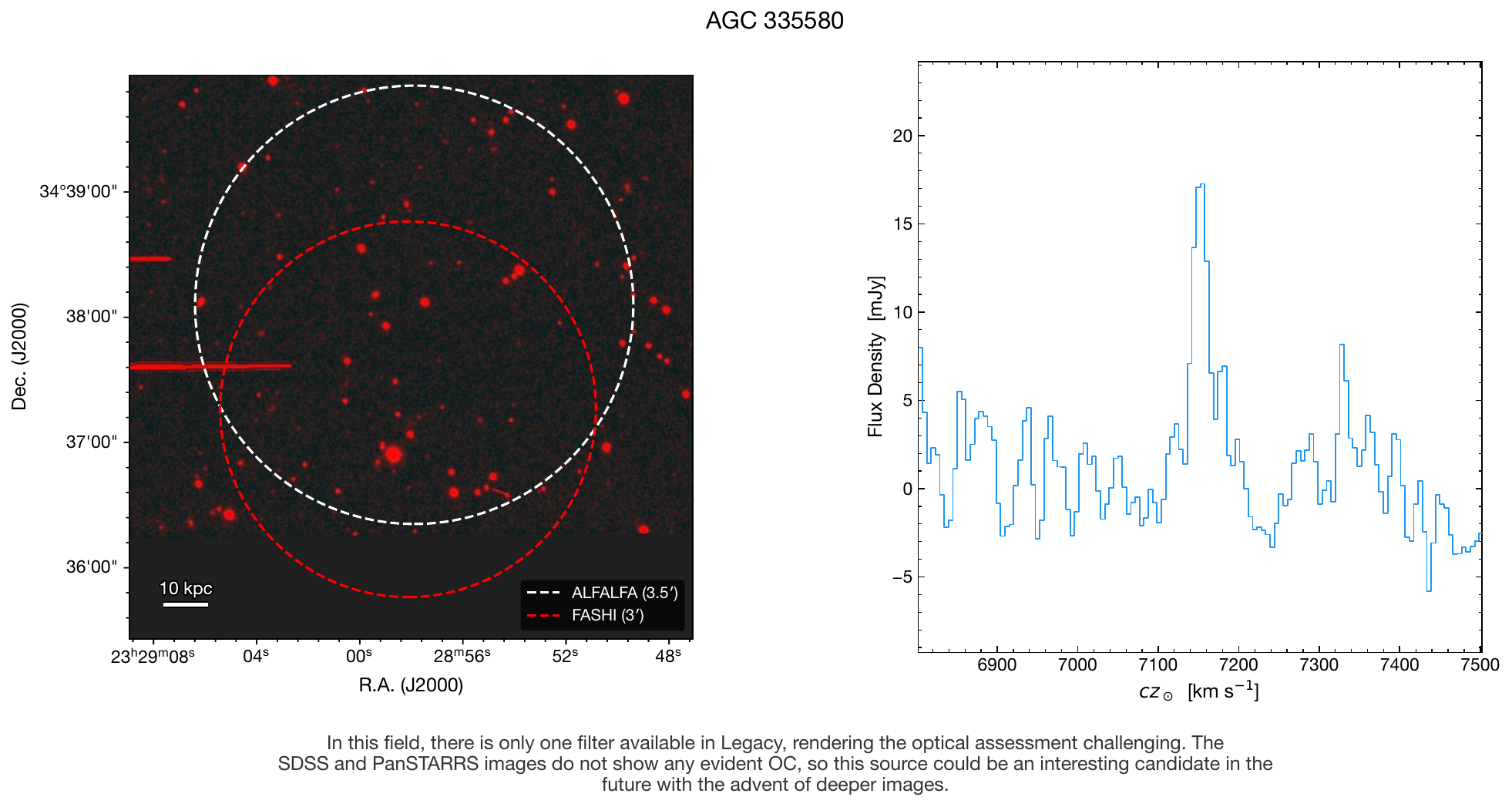}
\end{figure*}